\documentclass{osa-article}

\journal{oe}

\articletype{Research Article}

\usepackage{pgfplots}
\usetikzlibrary{pgfplots.groupplots, matrix}
\pgfplotsset{compat=1.15}
\usepackage[font=scriptsize, caption=false]{subfig}
\usepackage{graphicx}
\usepackage{multirow}
\usepackage{wrapfig}
\usepackage[section]{placeins}

\newtheorem{theorem}{Theorem}

\begin{document}
	
\title{Spatial Perspective Transform Estimation from Fourier Spectrum Analysis of 2D Patterns in 3D Space}

\author{Ian J. Maquignaz,\authormark{1,*}}

\address{Electrical and Computer Engineering,
Queen's University, Kingston, Ontario, Canada, K7L 3N6}
\email{\authormark{1,*}ian.maquignaz@queensu.ca}
RCV Lab \homepage{\authormark{1} https://rcvlab.engineering.queensu.ca/}
Ingenuity Labs \homepage{\authormark{1} https://ingenuitylabs.queensu.ca/}


\begin{abstract}
A novel approach to 3D surface imaging is proposed, allowing for the continuous sampling of 3D surfaces to extract localized perspective transformation coefficients from Fourier spectrum analysis of projected patterns. The mathematical relationship for Spatial-Fourier Transformation Pairs is derived, defining the transformation of spatial transformed planar surfaces in the Discrete Fourier Transform spectrum. The mathematical relationship for the twelve degrees of freedom in perspective transformation is defined and validated, asserting congruity with independent and uniform transform pairs for spatial Euclidean, similarity, affine and perspective transformations. This work expands on previously derived affine Spatial-Fourier Transformation Pairs and characterizes its implications towards 3D surface imaging as a means of augmenting (X,Y,Z)-(R,G,B) point-clouds to include additional information from localized sampling of pattern transformations.
\end{abstract}


\section{Introduction}

From optical communication to digital image processing, frequency domain transformations are a prominant tool used in optical signal analysis~\cite{paper:kikuchi}\cite[~p.249-354]{book:digimproc4ed}. In recent years, the research community has taken a growing interest in 3-dimensional (3D) surface imaging for different applications including range sensing, object/facial recognition, dynamic projection, 3D map building, and localization for autonomous vehicles. To make these applications possible, the analysis of structured light (SL) and coded structured light (CSL) for surface profilometry and 3D surface imaging have been proposed. As demonstrated by Salvi~et al.'s~\cite{paper:summaryOfStructuredLight} summary of works from 1982 to 2009 and Geng's~\cite{paper:tutorialSL} review of recent advances in surface imaging technology, this can be accomplished through a multitude of different approaches with a wide range of operating characteristics. For the purpose of this work, methods are categorized as either discrete or continuous active-correspondence, with sub-categorization as spatial, temporal, or spatio-temporal multiplexing. 

Discrete spatially multiplexed methodologies are prominent in consumer products through active-correspondence products such as the Microsoft Kinect v1~\cite{paper:kinect} and the Intel RealSense~\cite{web:RealSenseD435}. These products use infrared (IR) laser sources projected through diffraction gratings to project spatial patterns onto a scene which are visually imperceptible to a human observer and therefore unimpeding of an observer's viewpoint. Higher precision discrete methodologies have been proposed through spatio-temporal multiplexing, including recent work by Cole~et~al.~\cite{paper:avery} which proposed modulating imperceptible grey code patterns by controlling a DLP projector's digital mirroring device (DMD). This approach extends on previously reported spatio-temporal multiplexing of grey-codes by Cotting and Fuchs~\cite{paper:cotting}. Consumer examples of multi-pattern/multi-shot approaches are limited, likely indicative of the increased hardware costs associated with pattern modulation and/or the requirement for stationary subjects~\cite{web:zivid}. As described by Cole~et~al., a 2D-search requires $\log_2(M)+\log_2(N)$ grey-code patterns to produce a M$\times$N depth-map, thus requiring the projection of 22 unique patterns on a stationary subject for a 1920$\times$1080 depth-map.

Continuous patterns offer a higher tolerance for distortion and defocused optics which can be advantageous~\cite{paper:depthFromDefocus, paper:frequencyStructuredLightDepth}. The single phase shifting (SPS) methods by Srinivasana~et~al.~\cite{paper:Srinivasan} and Guan~et~al.~\cite{paper:Guan:03} employ sinusoidal gratings or fringe patterns to sense depth through phase shifts. Multiple phase shifting (MPS) methods extend these approaches through works such as Gushov~et~al.~\cite{paper:Gushov1991}, although these methods do not extend beyond the analysis of phase components for sinusoidal signals. The Fourier Transform becomes predominant with single coded frequency multiplexing (SCFM) approaches with the proposition of Fourier-Transform fringe-pattern analysis for topography and interferometry~\cite{paper:Takeda:82}, and the definition of Fourier Transform profilometry (FTP) in 1983 by Takeda~et~al.~\cite{paper:Takeda}. These early works were followed by Su~et~al.~\cite{paper:Su1990} who proposed the usage of a Ronchi grating to create a quasi-sine optical field which could be analyzed in the Fourier domain without overlapping the zero component and other higher spectra. Later works include a single-shot 3D shape measurement methodology using Frequency-multiplex Fourier-transform profilometry by Takeda~et~al.~\cite{paper:Takeda:97}. 

A review of FTP methodologies presented by Su and Chen in 2001~\cite{paper:SU2001263}, concluded that FTP methodologies offer great potential for further improvements, which was followed by a decade of Fourier contributions~\cite{paper:CHEN20051267,paper:YUE20071170,paper:GDEISAT2006482,paper:Wu2006,paper:CHEN2007821,paper:Berryman2008,paper:HU200957}. Among these works, Gong and Zhang proposed using sinusoidal fringe pattern for defocused binary patterns, achieving 480$\times$480 depth-maps at 4,000 Hz with believed potential to achieve 12,500 Hz by either lowering the projector resolution, or replacing it with a mechanical grating \cite{paper:Gong}. The predominant approach to Fourier analysis surface imaging is through phase shifting patterns, though some have explored higher dimensionality in Fourier transformations. Lin~and~Su~\cite{paper:LinSu} proposed using a two-dimensional (2D) FTP to provide better separation of depth information from speckle-like structure and discontinuity noise in fringe patterns, and Hallerman and Shirley~\cite{paper:Hallerman} proposed extracting height profiles from time-varying speckle-intensity patterns using a 3D Fast Fourier Transform, but examples of such endeavours are sparse. Within these works, the optical emitters and sensory hardware used to enable FTP methodologies varies, with approaches such as Gong and Zhang's using off-the-shelf consumer projectors~\cite{paper:Gong}, and others such as Hallerman and Shirly~\cite{paper:Hallerman} and Dresel~et al's~\cite{paper:Dresel:92} using laser sources. 

Recent approaches have evolved beyond pure depth estimation and introduced a new interest in pursuing a richer understanding of the properties of sensed surfaces. This includes works such as Jiang~et~al.~\cite{paper:Jiang:18} who proposed a two projector and one camera configuration to produce 3D shape measurements with reduced shadow by estimating the surface normal to reduce directional bias. Continuing with this initiative, the proposed method introduces a novel approach to pattern generation, embedding correspondences in the Fourier spectrum to enable the augmentation of point cloud data to include pattern transformation coefficients. This expansion on the capacities of Fourier-based surface imaging is possible through the definition of a mathematical framework for mapping spatial perspective transformations to the discrete Fourier spectrum. This work contrasts in approach from phase-oriented Fourier analysis prominent in previous works, and as will be demonstrated, offers new potential for CSL surface imaging.

The organization of the remainder of this paper is as follows: Section \ref{sec:PerspectiveTransformTheorem} introduces the Discrete Fourier Transformation and Perspective Transform Theorem as a means of deriving the proposed mathematical relationship between the spatial and Fourier domains as Spatial-Fourier Transformation Pairs (SFTP). Section~\ref{sec:experimentalResults} presents experimental results from the simulated transformation of images, validating the equations derived in Section~\ref{sec:PerspectiveTransformTheorem} by demonstrating congruence between the mathematical and simulated models. Section~\ref{sec:LimitationAndSpecialCases} extends on Section~\ref{sec:PerspectiveTransformTheorem} to define special cases of SFTP for translations and shearing onto the Z-axis (warp). The paper concludes with a summary and discussion in Section~\ref{sec:Discussion}.

\section{Spatial-Fourier Transformation Pairs}
\label{sec:PerspectiveTransformTheorem}

The proposed method allows for correspondences to be embedded and analyzed in the Fourier spectrum using a novel derivation rooted in the theory of Fourier and geometric transformation. For completeness in the derivation of the proposed perspective theorem for SFTP, some background in Fourier and transform theory is provided.

\subsection{The Fourier and Discrete Fourier Transform}
\label{subsec:FourierTransform}

The well known and centuries old Fourier Transform (FT; $\mathcal{F}$), as shown in its 3D form by Equation \ref{eq:ft_spatial_to_frequency}, samples an infinite 3D space $f(x,y,z)$ and produces the equivalent frequency domain representation $F(u,v,w)$. This is achieved by taking the input spatial signal and decomposing it into sine and cosine components such that the spatial signal can be represented exactly in the frequency domain as the sum of infinite sine and cosine functions. The inverse of the FT ($\mathcal{F}^{-1}$) shown in Equation \ref{eq:ft_frequency_to_spatial}, returns a frequency domain image to the spatial domain without loss.

\begin{equation}
F(u,v,w) = \mathcal{F}(f(x,y,z)) \Leftrightarrow f(x,y,z) = \mathcal{F}^{-1}(F(u,v,w))
\label{eq:ft_spatial_to_frequency_short}
\end{equation}
\begin{equation}
	F(u,v,w) = 
	\int_{-\infty}^{\infty}\int_{-\infty}^{\infty}\int_{-\infty}^{\infty}
	f(x,y,z)
	\textit{e}^{-2\imath\pi(ux+vy+wz)}\mathrm{d}x\mathrm{d}y\mathrm{d}z
\label{eq:ft_spatial_to_frequency}
\end{equation}

\begin{equation}
	f(x,y,z) = 
	\int_{-\infty}^{\infty}\int_{-\infty}^{\infty}\int_{-\infty}^{\infty}
	F(u,v,w)
	\textit{e}^{2\imath\pi(ux+vy+wz)}\mathrm{d}u\mathrm{d}v\mathrm{d}w
\label{eq:ft_frequency_to_spatial}
\end{equation}

Representing a continuous spatial image exactly in the frequency domain is impractical as the number of frequencies required to fully decompose an image is unbounded. To overcome this challenge, the Discrete Fourier Transform (DFT) can be used to impose boundaries on a FT, such that a discrete spatial signal can be represented in the frequency domain as the sum of a finite number of sampled sine and cosine functions. This is illustrated in Equation~\ref{eq:dft_frequency_to_spatial}, where the FT is bound by the number of sampled frequencies $M$ in the $x$-axis, $N$ in the $y$-axis, and $D$ in the $z$-axis, for a total of 
$M \times N \times D$ sampled frequencies. For consistency, throughout this work $M$ and $N$ are set respectively to the number of columns and rows in an input image. As will be discussed in the following section, for the purpose of this work $D$ is implicitly set to a value of 1 for 2D spatial input. 

\begin{equation}
	F(u,v,w) = 
	\sum_{x=0}^{M-1}
	\sum_{y=0}^{N-1}
	\sum_{z=0}^{D-1}
	f(x,y,z)
	\textit{e}^{-2\imath\pi\left(\frac{ux}{M}+\frac{vy}{N}+\frac{wz}{D}\right)}
\label{eq:dft_spatial_to_frequency}
\end{equation}

\begin{equation}
	f(x,y,z) = 
	\frac{1}{MND}
	\sum_{u=0}^{M-1}
	\sum_{v=0}^{N-1}
	\sum_{w=0}^{D-1}
	F(u,v,w)
	\textit{e}^{2\imath\pi\left(\frac{ux}{M}+\frac{vy}{N}+\frac{wz}{D}\right)}
\label{eq:dft_frequency_to_spatial}
\end{equation}

Limited derivations of the relationship between spatial transformations and the FT spectrum appear in the first (1977) and second edition (1987) of \textit{Digital Image Processing} by Gonzalez and Wintz~\cite[~p.47-78]{book:old_digimproc1ed}~\cite[~p.72-100]{book:old_digimproc2ed} and (1996) \textit{Digital Image Processing} by Castleman~\cite[~p.178-186]{book:DigitalImageProcessingCastleman}. The relationship between spatial affine transformations and the FT spectrum has been previously defined by Bracewell~et~al.\ in 1993~\cite{paper:BracewellAffine2dFFT}. Bracewell included derivations of the theorem for Euclidean, similarity, and affine transforms in \textit{Fourier Analysis and Imaging}~\cite[~p.154-161]{book:fourierimaging}. Derivations for Euclidean and similarity transformations appear in the second edition of \textit{Digital Image Processing} by Gonzalez and Woods~\cite[~p.210]{book:digimproc2ed}, but are subsequently omitted from the third and fourth editions of the same text~\cite{book:digimproc3ed,book:digimproc4ed}. In this work, the derivation extends Bracewell's approach and combines it with the observations and notational convention set forth by Brigham~\cite[~p.35]{book:fastFourierTransform}~\cite[~p.30-47]{book:fastFourierTransformAndApplications}.

\subsection{3D Perspective Transform of 2D Planes}
\label{subsec:PerspectiveTransformation}

As described by Hartley and Zisserman \cite{book:hartley2003multiple}, spatial transformations follow a well established hierarchy of classes, i.e. isometry, similarity, affinity, and perspectivity transformations. Each subsequent class in the hierarchy relaxes invariant properties, inheriting and expanding the degrees of freedom (DoFs) of previous classes. For example, affine transformations preserve parallel lines and represent six DoFs in 2D space, whereas perspective transformations preserves only collinearity and represents eight DoFs in 2D space~\cite[~p.16-22]{book:hartley2003multiple}. These DoFs increase with dimensionality increasing affine transformations to twelve DoF in 3D space and perspective transformations to fifteen DoFs~\cite[~p.59]{book:hartley2003multiple}. Equation~\ref{eq:perspective_transform} defines the transformation of a 2D point $P$ represented in homogeneous coordinates, using perspective transformation $A$ to produce point $P'$. The use of homogeneous coordinates adds the third dimension $z$, which is subsequently removed from $P'$ by perspective divide to produce $P''$ as shown in Equation~\ref{eq:perspective_divide}. 

\begin{equation}
\begin{bmatrix}
x' \\ y' \\ z'
\end{bmatrix}
=
P' = AP
= 
\begin{bmatrix}
{\chi}_x    && {\psi}_{yx} && {\tau}_{x}\\ 
{\psi}_{xy} && {\chi}_y    && {\tau}_{y}\\ 
{\psi}_{xz} && {\psi}_{yz} && {\chi}_z\\
\end{bmatrix}
\begin{bmatrix}
x \\ y \\ z=1
\end{bmatrix} 
\label{eq:perspective_transform}
\end{equation}

\begin{equation}
P'' = P' / z'
\label{eq:perspective_divide}
\end{equation}

The elements of matrix $A$ have been denoted with symbolic terms for scaling ($\chi$), shear ($\psi$) and translation ($\tau$) transformation coefficients. The scaling and translation elements are denoted as $\chi_{axis}$ and $\tau_{axis}$, where the subscript indicates the impacted axis. Shearing elements denote two axes in subscript, where $\chi_{ab}$ denotes the proportion of dimension $a$ to be added to dimension $b$. If expressed in terms of a planar image $f(x,y,1)$, Equations \ref{eq:perspective_transform} and \ref{eq:perspective_divide} can be rewritten as Equation \ref{eq:perspective_transform_full}, transforming $f(x,y,1)$ to planar image $f(x', y',1)$ which can be referred to independently as $g(x,y,1)$. 

\begin{equation}
g(x,y,1) = f(x',y',1)= f\left(\frac{\chi_xx + \psi_{yx}y + \tau_x}{\psi_{xz}x + \psi_{yz}y + \chi_zz}, \frac{\psi_{xy}x + \chi_yy + \tau_y}{\psi_{xz}x + \psi_{yz}y + \chi_zz}, 1\right)
\label{eq:perspective_transform_full}
\end{equation}

To simplify the relationship between spatial translation and phase in Fourier space, the transformation is decomposed to isolate translation. This is achieved by redistributing Equation~\ref{eq:perspective_transform} to produce Equation~\ref{eq:perspective_transform_decomposed}, where the perspective transform has been decomposed into matrix $B$ for shear, rotation and scaling, and a separate matrix $C$ for translation. The inverse of Equation~\ref{eq:perspective_transform_decomposed} is illustrated in Equation~\ref{eq:perspective_transform_inverse}.

\begin{equation}
P' = BP+C
= 
\begin{bmatrix}
{\chi}_x    && {\psi}_{yx} && 0\\ 
{\psi}_{xy} && {\chi}_y    && 0\\ 
{\psi}_{xz} && {\psi}_{yz} && {\chi}_z\\
\end{bmatrix}
\begin{bmatrix}
x \\ y \\ z=1
\end{bmatrix} 
+
\begin{bmatrix}
{\tau}_{x} \\ {\tau}_{y} \\ {\tau}_{z}=0
\end{bmatrix} 
\label{eq:perspective_transform_decomposed}
\end{equation}
\begin{equation}
\centering
P = B^{-1} (P'-C)
\label{eq:perspective_transform_inverse}
\end{equation}

\subsection{Derivation of Spatial-Fourier Transformation Theorem}
\label{subsec:PerspectiveTransformTheorem}

Following the precedent and convention set forth by Bracewell~et~al.~\cite{paper:BracewellAffine2dFFT} in deriving the 
\emph{Affine Theorem}, the \emph{Perspective Theorem} for SFTP is derived here by recognizing that if there exists a mapping between spatial image $f(x,y,z)$ and frequency domain representation $F(u,v,w)$, then given Equation \ref{eq:perspective_transform_full} for a perspective transformation and Equation \ref{eq:dft_spatial_to_frequency} for a DFT, a mapping from $f(x',y',z')$ to $F(u',v',w')$ must exist. 

\begin{theorem}[Perspective Transformation Theorem]
	\label{PTtheorem}
	If $f(x,y,z)$ has 3D \emph{DFT} 
	$F(u,v,w)$,
	then 
	$g(x,y,z) = f\left(\frac{\chi_xx + \psi_{yx}y + \tau_x}{\psi_{xz}x + \psi_{yz}y + \chi_zz}, \frac{\psi_{xy}x + \chi_yy + \tau_y}{\psi_{xz}x + \psi_{yz}y + \chi_zz}, 1\right)$
	has 3D \emph{DFT}
	$G(u, v, w)$, \\
	with $G(u, v, w) = EF(u,v,w)e^{2\pi\imath(EC)}$,
	and
	$E = \begin{bmatrix}
	\frac{u}{M} && \frac{v}{N} && \frac{w}{D}
	\end{bmatrix}
	B^{-1}$.
\end{theorem}

\noindent\emph{\textit{Proof.}}
To derive this result, the inner product of $[u\;v\;w]^\intercal$ and 
$[x\;y\;z]^\intercal$ is recognized in the phase exponent $-2\imath\pi(ux+vy+wz)$ of the 3D Fourier transform of Equation~\ref{eq:dft_spatial_to_frequency}. This inner product can be defined by Equation~\ref{eq:innerProduct_dft} for a spatial point ${P_s}$ in an $M\times N$ image with homogeneous 3\textsuperscript{rd} dimension $D$. 

\begin{equation}
\frac{ux}{M} + \frac{vy}{N} + \frac{wz}{D}=
\begin{bmatrix}
\frac{u}{M} && \frac{v}{N} && \frac{w}{D}
\end{bmatrix}
P_s
\label{eq:innerProduct_dft}
\end{equation}

\begin{equation}
\frac{ux}{M} + \frac{vy}{N} + \frac{wz}{D}=
\begin{bmatrix}
\frac{u}{M} && \frac{v}{N} && \frac{w}{D}
\end{bmatrix}
B^{-1}(P'_s-C)
\label{eq:innerProduct_XYZ_UVW_expansion}
\end{equation}

\noindent The perspective transformation coefficients can be substituted into Equation~\ref{eq:innerProduct_dft} to produce Equation~\ref{eq:innerProduct_XYZ_UVW_expansion}. This substitution is possible by taking the inverse of the perspective transformation as shown in Equation~\ref{eq:perspective_transform_inverse}. 

\begin{equation}
\frac{ux}{M} + \frac{vy}{N} + \frac{wz}{D}=E(P'_s-C)\text{, where }E= \begin{bmatrix}
\frac{u}{M} && \frac{v}{N} && \frac{w}{D}
\end{bmatrix}
B^{-1}
\label{eq:convolution_XYZ_UVW}
\end{equation}

\noindent Reducing Equation \ref{eq:innerProduct_XYZ_UVW_expansion} into Equation \ref{eq:convolution_XYZ_UVW}, components can then be re-substituted back into Equation \ref{eq:dft_spatial_to_frequency} to produce Equation \ref{eq:DFT_3D_intermediate}, mapping spatial image $f(x',y',z')$ to frequency representation $F(u',v',w')$. 

\begin{equation}
F(u',v',w') = 
\sum_{x=0}^{M-1}  \sum_{y=0}^{N-1} \sum_{z=0}^{D-1} 
f(x', y', z') e^{-2\pi\imath(EP'_s)} e^{2\pi\imath(EC)}
\label{eq:DFT_3D_intermediate}
\end{equation}

\noindent This completes the derivation of the theorem, proving the existence of a mapping for 2D spatial perspective transformations between frequency domain images $F(u,v,w)$ and $F(u',v',w')$. \\

Following the precedent of Equation~\ref{eq:perspective_transform_full}, the spatial perspective transformation of a frequency domain image $F(u,v,w)$ to $F(u',v',w')$ can be expressed as Equation~\ref{eq:dft_3D_transformed_image}. This reduction is possible by substituting $F(u,v,w)$ for the Fourier Transform component expressed in Equation~\ref{eq:dft_spatial_to_frequency}, re-writing periodicity terms in $E$ as shown in Equation~\ref{eq:convolution_XYZ_UVW_NO_DIMENTIONALITY} (see Section~\ref{subsec:translation}) and retaining only spatial perspective transformation coefficients $H$ and $e^{2\imath\pi(EC)}$, which respectively represent a transformation of frequency space and a phase-shifting component. To simplify, the mapping can be expressed as Equation~\ref{eq:dft_3D_transformed}, which provides the mapping in terms frequency points $P_f$ and $P_f'$, as previously demonstrated for spatial points $P_s$ and $P'_s$ in Equation~\ref{eq:perspective_transform}. 

\begin{equation}
H = \begin{bmatrix}
{u} && {v} && {w}
\end{bmatrix}
B^{-1}
\label{eq:convolution_XYZ_UVW_NO_DIMENTIONALITY}
\end{equation}

\begin{equation}
F(u', v', w')
=
HF(u,v,w)e^{2\pi\imath(EC)}
\label{eq:dft_3D_transformed_image}
\end{equation}
\begin{equation}
P_f'
=
HP_fe^{2\pi\imath(EC)} \text{, where }P_f=\begin{bmatrix}
u && v && w
\end{bmatrix}^T
\label{eq:dft_3D_transformed}
\end{equation}

\subsection{Perspective Transformation Pairs}

To increase understanding and utility of SFTP, short derivations of Equation~\ref{eq:dft_3D_transformed} can be used as transformation pairs, defining mappings of Fourier space for simple spatial transformations. Such pairs are possible by recognizing the relationship between spatial and Fourier space as shown in Equation~\ref{eq:ft_spatial_to_frequency_short} and redefining in terms of pairwise relationships as shown in Equation~\ref{eq:ftp_pair}.
	
\begin{equation}
f(x,y) \Leftrightarrow F(u,v)
\label{eq:ftp_pair}
\end{equation}
\begin{equation}
|F(u,v)| = \sqrt{R(u,v)^{2} + I(u,v)^{2}}
\label{eq:ft_magnitude}
\end{equation}

Taking the magnitude of the Fourier spectrum using Equation~\ref{eq:ft_magnitude}, the real and imaginary spectra of the Fourier transform are concatenated to produce a single representation invariant of spatial translation. This allows for the isolation of perspective transformation coefficients and the simplification of Equation~\ref{eq:dft_3D_transformed} to Equation~\ref{eq:dft_3D_transformed_Mag}:

\begin{equation}
P_F'= HP_F
\label{eq:dft_3D_transformed_Mag}
\end{equation}

\noindent Using Equation~\ref{eq:dft_3D_transformed_Mag}, transform pairs can be derived for a subset of isometric, similarity, affine and perspective transformations. Table~\ref{table:Transform_Pairs} summarizes a number of these derivations. Translation and shearing onto the Z-axis (warp) represent special cases and are described in Section~\ref{sec:LimitationAndSpecialCases}.

\begin{table}[ht]
	\centering
	\caption{2D Transformation pairs}
	\begin{tabular}{|ll|}
		\hline
		\textit{Rotation:} &
		$f(\cos({\theta})x+\sin({\theta})y,\cos({\theta})y-\sin({\theta})x,1)$
		\\
		& \hspace{0.5 cm} $\Leftrightarrow 
		F\left(
		{\cos({\theta})u \; - \; \sin({\theta})v},
		{\cos({\theta})v \; + \; \sin({\theta})u},
		1\right)$                                        \\ \hline
		\textit{Scaling:} &
		$f({\chi}_{x}x,{\chi}_{y}y, 1)\Leftrightarrow F(\frac{u}{{\chi}_{x}},\frac{v}{{\chi}_{y}},1)$
		\\
		&	$f({\chi}_{x}x,{\chi}_{y}y,{\chi}_{z}z) \Leftrightarrow F(\frac{u}{{\chi}_{x}},\frac{v}{{\chi}_{y}},\frac{w}{{\chi}_{z}})$
		\\ \hline
		\textit{Shearing:} & 
		$f(x+{\psi}_{yx}y,y+{\psi}_{xy}x,1) \Leftrightarrow F\left(\frac{u-{\psi}_{xy}v}{1-{\psi}_{yx}{\psi}_{xy}},\frac{v-{\psi}_{yx}u}{1-{\psi}_{yx}{\psi}_{xy}},1\right)$                                        \\ \hline
	\end{tabular}
	\label{table:Transform_Pairs}
\end{table}

\section{Experimental Validation}
\label{sec:experimentalResults}

To validate Theorem~\ref{PTtheorem}, a experiment was designed to emulate the operation of a Projector-Camera (PROCAM) pair in a real-world environment. In this configuration, the projection and subsequent imaging of 2D patterns in 3D space was emulated, providing validation of the derived SFTP and confirmation of its applicability towards the proposed application of surface sensing. To achieve this, images were encoded with known frequency domain patterns and subjected to known transformations in the spatial domain, allowing for empirical measurement of conformance between Equation~\ref{eq:dft_3D_transformed_Mag} and transformations measured from imaging. 

\begin{figure}[h]
	\centering
	\subfloat[Image $I$]{
		\includegraphics[width=0.22\textwidth]{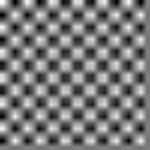}
		\label{fig:NO_TRANSFORM_rawSB}
	}
	\subfloat[$Re(\mathcal{F}(I))$]{
		\includegraphics[width=0.22\textwidth]{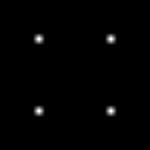}
		\label{fig:NO_TRANSFORM_rawF_SB__Re}
	}
	\subfloat[$Im(\mathcal{F}(I))$]{
		\includegraphics[width=0.22\textwidth]{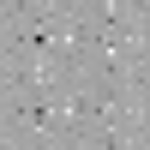}
		\label{fig:NO_TRANSFORM_rawF_SB__Im}
	}
	\subfloat[$|\mathcal{F}(I)|$]{
		\includegraphics[width=0.22\textwidth]{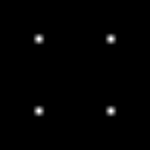}
		\label{fig:NO_TRANSFORM_rawF(SB)}
	}
	\caption{A sample set of unaltered encoded images used for validation.}
	\label{figs:NO_TRANSFORM}
\end{figure}

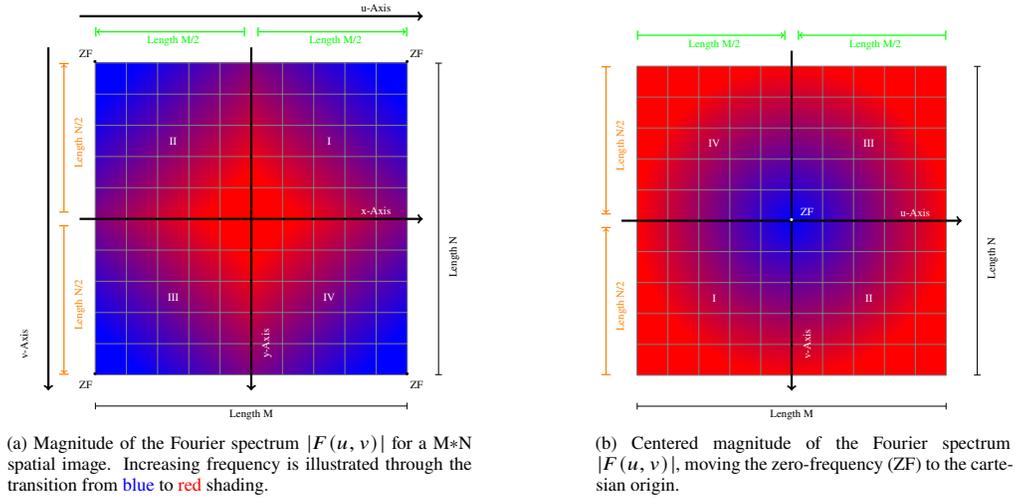
\begin{figure}[!ht]
	\centering
	\subfloat[Magnitude of the Fourier spectrum $|F(u,v)|$ for a M$*$N spatial image. Increasing frequency is illustrated through the transition from \textcolor{blue}{blue} to \textcolor{red}{red} shading.]{
		\resizebox {0.45\textwidth} {!} {
			\begin{tikzpicture}
			\shade[top color=blue, right color=red, shading angle=-45] (0,0) rectangle (5,5);
			\shade[top color=blue, right color=red, shading angle=45] (0,0) rectangle (-5,5);
			\shade[top color=red, right color=blue, shading angle=-45] (0,0) rectangle (-5,-5);
			\shade[top color=red, right color=blue, shading angle=45] (0,0) rectangle (5,-5);
			
			\draw[step=1cm,gray,very thin] (-5,-5) grid (5,5);
			\node[xshift=2.5cm, yshift=2.5cm, color=white]{I};
			\node[xshift=-2.5cm, yshift=2.5cm, color=white]{II};
			\node[xshift=-2.5cm, yshift=-2.5cm, color=white]{III};
			\node[xshift=2.5cm, yshift=-2.5cm, color=white]{IV};
			
			\node[xshift=-5cm, yshift=-5cm]{\textbullet};
			\node[xshift=5cm, yshift=-5cm]{\textbullet};
			\node[xshift=-5cm, yshift=5cm]{\textbullet};
			\node[xshift=5cm, yshift=5cm]{\textbullet};
			\node[xshift=-5.3cm, yshift=-5.3cm]{ZF};
			\node[xshift=5.3cm, yshift=-5.3cm]{ZF};
			\node[xshift=-5.3cm, yshift=5.3cm]{ZF};
			\node[xshift=5.3cm, yshift=5.3cm]{ZF};
			
			\draw[line width = 0.7mm,->] (-5.5,6.5) -- (5.5,6.5) node[above,xshift=-1.5cm, color=black]{u-Axis};
			
			\draw[line width = 0.7mm,->] (-6.5,5.5) -- (-6.5,-5.5) node[above,xshift=-0.5cm,yshift=1.5cm,rotate=90, color=black]{v-Axis};
			
			\draw[line width = 0.7mm,->] (-5.5,0) -- (5.5,0) node[above,xshift=-1.5cm, color=white]{\textcolor{white}{x-Axis}};
			
			\draw[line width = 0.7mm,<-] (0,-5.5) -- (0,5.5) node[right,xshift=0.5cm,yshift=-10.1cm,rotate=90, color=white]{\textcolor{white}{y-Axis}};

			\draw[|-|] (-5,-6) -- (5,-6) node[below,xshift=-5cm]{Length M};
			\draw[|-|] (6,-5) -- (6,5) node[right,xshift=0.5cm,yshift=-7cm,rotate=90]{Length N};
			
			\draw[line width = 0.4mm,|<-|,color=green] (-5,6) -- (-0.2,6) node[below,xshift=-2.3cm]{Length M/2};
			\draw[line width = 0.4mm,|->|,color=green] (0.2,6) -- (5,6) node[below,xshift=-2.3cm]{Length M/2};
			\draw[line width = 0.4mm,|<-|,color=orange ] (-6,-5) -- (-6,-0.2) node[right,xshift=0.5cm,yshift=-3.5cm,rotate=90]{Length N/2};
			\draw[line width = 0.4mm,|->|,color=orange] (-6,0.2) -- (-6,5) node[right,xshift=0.5cm,yshift=-3.5cm,rotate=90]{Length N/2};
			\end{tikzpicture}
			\label{drawing:fft}
		}
	} 
	\hfill
	\subfloat[Centered magnitude of the Fourier spectrum $|F(u,v)|$, moving the zero-frequency (ZF) to the cartesian origin.]{
		\resizebox {0.4\textwidth} {!} {
			\begin{tikzpicture}
			\shade[outer color=red,inner color=blue] (-5,-5) rectangle (5,5);
			
			\draw[step=1cm,gray,very thin] (-5,-5) grid (5,5);
			\node[xshift=2.5cm, yshift=2.5cm, color=white]{III};
			\node[xshift=-2.5cm, yshift=2.5cm, color=white]{IV};
			\node[xshift=-2.5cm, yshift=-2.5cm, color=white]{I};
			\node[xshift=2.5cm, yshift=-2.5cm, color=white]{II};
			
			\draw[line width = 0.7mm,->] (-5.5,0) -- (5.5,0) node[above,xshift=-1.5cm, color=white]{\textcolor{white}{u-Axis}};
			\draw[line width = 0.7mm,<-] (0,-5.5) -- (0,5.5) node[right,xshift=0.5cm,yshift=-10.1cm,rotate=90, color=white]{\textcolor{white}{v-Axis}};
			
			\node[xshift=0cm, yshift=0cm, color=white]{\textbullet};
			\node[xshift=0.5cm, yshift=0.3cm, color=white]{ZF};
			
			\draw[|-|] (-5,-6) -- (5,-6) node[below,xshift=-5cm]{Length M};
			\draw[|-|] (6,-5) -- (6,5) node[right,xshift=0.5cm,yshift=-7cm,rotate=90]{Length N};
			
			\draw[line width = 0.4mm,|->|,color=green] (-5,6) -- (-0.2,6) node[below,xshift=-2.3cm]{Length M/2};
			\draw[line width = 0.4mm,|<-|,color=green] (0.2,6) -- (5,6) node[below,xshift=-2.3cm]{Length M/2};
			\draw[line width = 0.4mm,|->|,color=orange] (-6,-5) -- (-6,-0.2) node[right,xshift=0.5cm,yshift=-3.5cm,rotate=90]{Length N/2};
			\draw[line width = 0.4mm,|<-|,color=orange] (-6,0.2) -- (-6,5) node[right,xshift=0.5cm,yshift=-3.5cm,rotate=90]{Length N/2};	
			\end{tikzpicture}
			\label{drawing:fft_centered}
		}
	}
	\caption{The DFT spectrum (\ref{drawing:fft}) and Centered DFT spectrum (\ref{drawing:fft_centered})}
	\label{figs:DFT_spectrum}
\end{figure}

As a base, an image of size 25$\times$25 was selected and encoded in the frequency domain with four frequencies of amplitude 10,000 at Cartesian coordinates $(6,6)$,$(-6,-6)$,$(6,-6)$, and $(-6,6)$ of the Fourier spectrum $F(u,v)$. To accommodate the use of Cartesian coordinates, the Fourier spectrum was reshaped to Cartesian representation using Equation~\ref{eq:FT_centeredPair}, centering the Fourier zero-frequency (ZF) to the Cartesian origin (0,0). This reshaping and its impact on the interpretation of the Fourier spectrum can be visualized through the transformation of Figure~\ref{drawing:fft} to Figure~\ref{drawing:fft_centered}. The resulting spatial image and its Fourier decomposition are illustrated in Figure~\ref{figs:NO_TRANSFORM}, with its spatial representation shown in Figure~\ref{fig:NO_TRANSFORM_rawSB}, real component of the Fourier spectrum in Figure~\ref{fig:NO_TRANSFORM_rawF_SB__Re}, imaginary component in Figure~\ref{fig:NO_TRANSFORM_rawF_SB__Im}, and magnitude in Figure~\ref{fig:NO_TRANSFORM_rawF(SB)}. To allow visualization of the spatial and Fourier spectrums, spatial transformation employed bilinear interpolation and all images represented in this works are normalized by Equation \ref{eq:NORM_MINMAX} to redistribute image minimum $f_{MIN}$ and maxima $f_{MAX}$ of the spectrum values to a uniform range of $MIN=0$ to $MAX=255$. To increase understanding of the Fourier spectrum, images of the frequency domain have the ZF zeroed to facilitate normalization.

\begin{equation}
\centering
f(x,y)(-1)e^{(x+y)}
\Leftrightarrow 
F(u-\frac{M}{2},v-\frac{N}{2})
\label{eq:FT_centeredPair}
\end{equation}

\begin{equation}
\centering
||f(x,y)|| =  {MIN} + \frac{(f(x,y)-f_{MIN})*({MAX}-{MIN})}{f_{MAX}-f_{MIN}}
\label{eq:NORM_MINMAX}
\end{equation}

\begin{figure}[ht]
	\scriptsize
	\centering
	\subfloat[${\chi}_{x}{=}1.25$]{
		\includegraphics[width=0.19\textwidth]{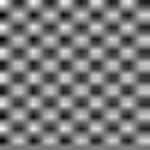}
		\label{fig:Scale_X_by2_rawSB}
	}
	\subfloat[${\chi}_{y}{=}1.25$]{
		\includegraphics[width=0.19\textwidth]{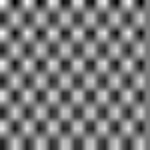}
		\label{fig:Scale_Y_by2_rawSB}
	}
	\subfloat[${\chi}_{z}{=}0.75$]{
		\includegraphics[width=0.19\textwidth]{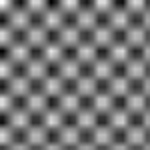}
		\label{fig:Scale_Z_by0_5_rawSB}
	}
	\subfloat[${\chi}_{x}{=}{\chi}_{y}{=}0.8$]{
		\includegraphics[width=0.19\textwidth]{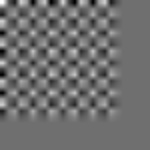}
		\label{fig:Scale_XY_by0_8_rawSB}
	}
	\subfloat[${\chi}_{x}{=}{\chi}_{y}{=}{\chi}_{z}{=}0.75$]{
		\includegraphics[width=0.19\textwidth]{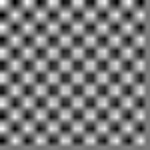} 
		\label{fig:Scale_XY_by0_5_rawSB}
	}
	\hfill
	\subfloat[$|\mathcal{F}({\chi}_{x}{=}1.25)|$]{
		\includegraphics[width=0.19\textwidth]{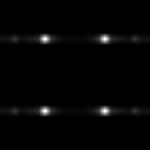}
		\label{fig:Scale_X_by2_rawF(SB)}
	}
	\subfloat[$|\mathcal{F}({\chi}_{y}{=}1.25)|$]{
		\includegraphics[width=0.19\textwidth]{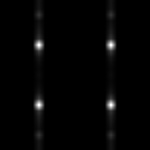}
		\label{fig:Scale_Y_by2_rawF(SB)}
	}
	\subfloat[$|\mathcal{F}({\chi}_{z}{=}0.75)|$]{
		\includegraphics[width=0.19\textwidth]{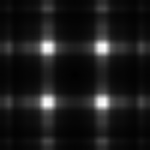}
		\label{fig:Scale_Z_by0_5_rawF(SB)}
	}
	\subfloat[$|\mathcal{F}({\chi}_{x}{=}{\chi}_{y}{=}0.8)|$]{
		\includegraphics[width=0.19\textwidth]{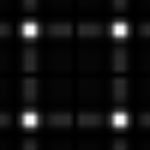}
		\label{fig:Scale_XY_by0_8_rawF(SB)}
	}
	\subfloat[$|\mathcal{F}({\chi}_{x}{=}{\chi}_{y}{=}{\chi}_{z}{=}0.75)|$]{
		\includegraphics[width=0.19\textwidth]{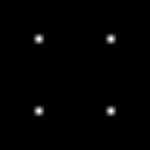}
		\label{fig:Scale_XYZ_by0_5_rawF(SB)}
	}
	\caption{Independent and uniform spatial scaling transformations of the $x$, $y$ \& $z$-axes}
	\label{figs:Scale_XYZ}
\end{figure}

\begin{figure}[ht]
	\centering
	\subfloat[${\psi}_{yx}{=}0.3$]{
		\includegraphics[width=0.19\textwidth]{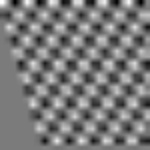}
		\label{fig:Shear_X_by0_3_rawSB}
	}
	\subfloat[${\psi}_{xy}{=}0.3$]{
		\includegraphics[width=0.19\textwidth]{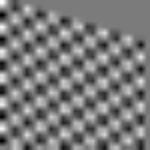}
		\label{fig:Shear_Y_by0_3_rawSB}
	}
	\subfloat[${\psi}_{yx}{=}{\psi}_{xy}{=}0.2$]{
		\includegraphics[width=0.19\textwidth]{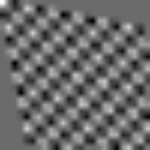}
		\label{fig:Shear_XY_by0_2_rawSB}
	}
	\subfloat[Rot(30$^\circ$)]{
		\includegraphics[width=0.19\textwidth]{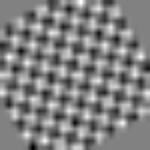}
		\label{fig:Rotation30_rawSB}
	}
	\subfloat[Rot(60$^\circ$)]{
		\includegraphics[width=0.19\textwidth]{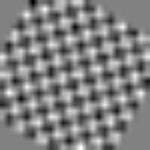}
		\label{fig:Rotation60_rawSB}
	}
	\hfill
	\subfloat[$|\mathcal{F}({\psi}_{yx}{=}0.3)|$]{
		\includegraphics[width=0.19\textwidth]{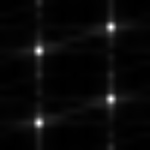}
		\label{fig:Shear_X_by0_3_rawF(SB)}
	}
	\subfloat[$|\mathcal{F}({\psi}_{xy}{=}0.3)|$]{
		\includegraphics[width=0.19\textwidth]{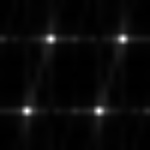}
		\label{fig:Shear_Y_by0_3_rawF(SB)}
	}
	\subfloat[$|\mathcal{F}({\psi}_{yx}{=}{\psi}_{xy}{=}0.2)|$]{
		\includegraphics[width=0.19\textwidth]{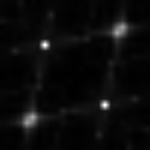}
		\label{fig:Shear_XY_by0_2_rawF(SB)}
	}
	\subfloat[$|\mathcal{F}($Rot(30$^\circ$)$)|$]{
		\includegraphics[width=0.19\textwidth]{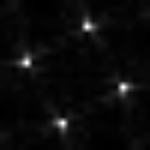}
		\label{fig:Rotation30_rawF(SB)}
	}
	\subfloat[$|\mathcal{F}($Rot(60$^\circ$)$)|$]{
		\includegraphics[width=0.19\textwidth]{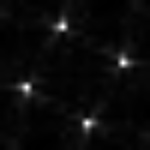}
		\label{fig:Rotation60_rawF(SB)}
	}
	
	\caption{Independent and uniform spatial shearing and rotation transformations of the $x$ \& $y$-axes}
	\label{figs:ShearXY}
\end{figure}

Figure~\ref{figs:Scale_XYZ} illustrates five samples taken to validate independent and uniform scaling of the $x$, $y$ and $z$-axes. To facilitate interpretation, Table~\ref{table:ScaleXYZ} summarizes the estimated and captured coordinates for each of the images in Figure~\ref{figs:Scale_XYZ}, with transformed coordinates identified as the four highest normalized cross-correlated template matched values in $|F(u,v)|$. This illustrates the conformity between the estimated and experimental results, with minor rounding resulting from integer pixel coordinates. This conformity is mirrored for shearing transformations as illustrated in Figure~\ref{figs:ShearXY} and summarized in Table~\ref{table:ShearXY}.

\begin{table}[ht]
	\centering
	\scriptsize
	\caption{Independent and uniform spatial scaling transformations of the $x$, $y$ \& $z$-axes}
	\begin{tabular}{c|c|c|c}
		\hline
		\textbf{Transform}	&  \textbf{Captured Points} & \textbf{Calculated Points} & \textbf{Estimated Transform} \\ \hline
		${\chi}_{x}=1.25$               	
		& \begin{tabular}[c]{@{}c@{}}(5, 6)(-5, -6)(5, -6)(-5, 6)\end{tabular}
		& \begin{tabular}[c]{@{}c@{}}(5, 6)(-5, -6)(5, -6)(-5, 6)\end{tabular} 
		& ${\chi}_{x}=1.25$
		\\ \hline
		${\chi}_{y}=1.25$ 
		& \begin{tabular}[c]{@{}c@{}}(6, 5)(-6, -5)(6, -5)(-6, 5)\end{tabular}
		& \begin{tabular}[c]{@{}c@{}}(6, 5)(-6, -5)(6, -5)(-6, 5)\end{tabular}
		& ${\chi}_{y}=1.25$
		\\ \hline
		${\chi}_{z}=0.75$ 	
		& \begin{tabular}[c]{@{}c@{}}(5, 5)(-5, -5)(5, -5)(-5, 5)\end{tabular}
		& \begin{tabular}[c]{@{}c@{}}(5, 5)(-5, -5)(5, -5)(-5, 5)\end{tabular}
		& ${\chi}_{z}=0.75$ 
		\\ \hline
		${\chi}_{x}={\chi}_{y}=0.8$  	
		& \begin{tabular}[c]{@{}c@{}}(7, 7)(-7, -7)(7, -7)(-7, 7)\end{tabular}
		& \begin{tabular}[c]{@{}c@{}}(7.5, 7.5)(-7.5, -7.5)(7.5, -7.5)(-7.5, 7.5)\end{tabular}
		& ${\chi}_{x}={\chi}_{y}=0.8$ 
		\\ \hline
		${\chi}_{x}={\chi}_{y}={\chi}_{z}=0.75$ 
		& \begin{tabular}[c]{@{}c@{}}(6, 6)(-6, -6)(6, -6)(-6, 6)\end{tabular} 
		& \begin{tabular}[c]{@{}c@{}}(6, 6)(-6, -6)(6, -6)(-6, 6)\end{tabular}
		& ${\chi}_{x}={\chi}_{y}={\chi}_{z}=0.75$
		\\ \hline	 
	\end{tabular}
	\label{table:ScaleXYZ}
\end{table}
\begin{table}[!ht]
	\centering
	\scriptsize
	\caption{Independent and uniform spatial shearing and rotation transformations of the $x$ \& $y$-axes}
	\begin{tabular}{c|c|c|c}
		\hline
		\textbf{Transform} & \textbf{Captured Points} & \textbf{Calculated Points} & \textbf{Estimated Transform} \\ \hline
		${\psi}_{yx}=0.3$
		& \begin{tabular}[c]{@{}c@{}}(6, 8)(-6, -8)(6, -4)(-6, 4)\end{tabular}
		& \begin{tabular}[c]{@{}c@{}}(6, 7.8)(-6, -7.8)(6, -4.2)(-6, 4.2)\end{tabular}
		& ${\psi}_{yx}=0.3$
		\\ \hline
		${\psi}_{xy}=0.3$	
		& \begin{tabular}[c]{@{}c@{}}(8, 6)(-8, -6)(4, -6)(-4, 6)\end{tabular}
		& \begin{tabular}[c]{@{}c@{}}(7.8, 6)(-7.8, -6)(4.2, -6)(-4.2, 6)\end{tabular}
		& ${\psi}_{xy}=0.3$
		\\ \hline
		${\psi}_{yx}={\psi}_{xy}=0.2$	
		& \begin{tabular}[c]{@{}c@{}}(7, 7)(-7, -7)(5, -5)(-5,5)\end{tabular}
		& \begin{tabular}[c]{@{}c@{}}(7.5, 7.5)(-7.5, -7.5)(5, -5)(-5,5)\end{tabular}
		& ${\psi}_{yx}={\psi}_{xy}=0.2$
		\\ \hline
		30$^\circ$ Rotation	
		& \begin{tabular}[c]{@{}c@{}}(2, 8)(-2, -8)(8, -2)(-8, 2)\end{tabular} 
		& \begin{tabular}[c]{@{}c@{}}(2.196, 8.196)(-2.196, -8.196)\\(8.196, -2.196)(-8.196, 2.196)\end{tabular}
		& 30$^\circ$ Rotation	
		\\ \hline	
		60$^\circ$ Rotation	
		& \begin{tabular}[c]{@{}c@{}}(8, 2)(-8, -2)(2, -8)(-2, 8)\end{tabular} 
		& \begin{tabular}[c]{@{}c@{}}(8.196, 2.196)(-8.196, -2.196)\\(2.196, -8.196)(-2.196, 8.196)\end{tabular}
		& 60$^\circ$ Rotation
		\\ \hline		
	\end{tabular}
	\label{table:ShearXY}
\end{table}
\begin{figure}[!ht]
	\centering
	
	\resizebox{\textwidth}{!} {
		\begin{tikzpicture}
		\begin{groupplot}[
		group style={
			group name=myPlots,
			group size=5 by 2,
			vertical sep=16pt, 
			horizontal sep=16pt,
		},
		footnotesize,
		xmin=0.7, xmax=1.3,
		ymin=4.5, ymax=8.5,
		xtick={0.0,0.2,...,4.0},
		ytick={0.0,1.0,...,9.0},
		minor x tick num=100,
		minor y tick num=1,
		]
		
		\nextgroupplot[xticklabels=\empty]
		\addplot [style={solid}, mark=none, color=red] table [col sep=comma, x={Scale_X}, y={Mag_X[6 6]}]{Scale_X.csv};
		\addlegendentry{$\chi_{x}$}
		
		\nextgroupplot[xticklabels=\empty, yticklabels=\empty]
		\addplot [style={solid}, mark=none, color=red] table [col sep=comma, x={Scale_XY}, y={Mag_X[6 6]}]{Scale_XY.csv};
		\addlegendentry{$\chi_{x}=\chi_{y}$}
		
		\nextgroupplot[xticklabels=\empty, yticklabels=\empty]
		\addplot [style={solid}, mark=none, color=red] table [col sep=comma, x={Scale_Z}, y={Mag_X[6 6]}]{Scale_Z.csv};
		\addlegendentry{$\chi_{z}$}
		
		\nextgroupplot[xticklabels=\empty, xmin=-0.05, xmax=0.35, ymin=5.5, ymax=9.5]
		\addplot [style={solid}, mark=none, color=red] table [col sep=comma, x={Shear_Y}, y={Mag_X[6 6]}]{Shear_Y.csv};
		\addlegendentry{$\psi_{xy}$}
		
		\nextgroupplot[xticklabels=\empty, yticklabels=\empty, xmin=-0.05, xmax=0.4, ymin=5.5, ymax=9.5]
		\addplot [style={solid}, mark=none, color=red] table [col sep=comma, x={Shear_XY}, y={Mag_X[6 6]}]{Shear_XY.csv};
		\addlegendentry{$\psi_{yx}=\psi_{xy}$}
		
		
		\nextgroupplot
		\addplot [style={solid}, mark=none, color=red] table [col sep=comma, x={Scale_X}, y={Est_Xi[6 6]}]{Scale_X.csv};
		\addplot [style={dotted}, mark=none, color=red] table [col sep=comma, x={Scale_X}, y={Est_Xd[6 6]}]{Scale_X.csv};
		\addlegendentry{$\text{Est. } \chi_{x}$}
		
		\nextgroupplot[yticklabels=\empty]
		\addplot [style={solid}, mark=none, color=red] table [col sep=comma, x={Scale_XY}, y={Est_Xi[6 6]}]{Scale_XY.csv};
		\addplot [style={dotted}, mark=none, color=red] table [col sep=comma, x={Scale_XY}, y={Est_Xd[6 6]}]{Scale_XY.csv};
		\addlegendentry{$\text{Est. } \chi_{x}=\chi_{y}$}	
		
		\nextgroupplot[yticklabels=\empty]
		\addplot [style={solid}, mark=none, color=red] table [col sep=comma, x={Scale_Z}, y={Est_Xi[6 6]}]{Scale_Z.csv};
		\addplot [style={dotted}, mark=none, color=red] table [col sep=comma, x={Scale_Z}, y={Est_Xd[6 6]}]{Scale_Z.csv};
		\addlegendentry{$\text{Est. } \chi_{z}$}
		
		\nextgroupplot[xmin=-0.05, xmax=0.35, ymin=5.5, ymax=9.5]
		\addplot [style={solid}, mark=none, color=red] table [col sep=comma, x={Shear_Y}, y={Est_Xi[6 6]}]{Shear_Y.csv};
		\addplot [style={dotted}, mark=none, color=red] table [col sep=comma, x={Shear_Y}, y={Est_Xd[6 6]}]{Shear_Y.csv};
		\addlegendentry{$\text{Est. } \psi_{xy}$}
		
		\nextgroupplot[yticklabels=\empty, xmin=-0.05, xmax=0.4, ymin=5.5, ymax=9.5]
		\addplot [style={solid}, mark=none, color=red] table [col sep=comma, x={Shear_XY}, y={Est_Xi[6 6]}]{Shear_XY.csv};
		\addplot [style={dotted}, mark=none, color=red] table [col sep=comma, x={Shear_XY}, y={Est_Xd[6 6]}]{Shear_XY.csv};
		\addlegendentry{$\text{Est. } \psi_{yx}=\psi_{xy}$}
		
		\end{groupplot}
		
		\path [nodes={anchor=south,rotate=90,midway}]
		(myPlots c1r1.outer north west)--(myPlots c1r2.outer south west)
		node {$|F(u',v',z')|$ x-coordinate}
		;
		\path [nodes={anchor=north,midway}]
		(myPlots c1r2.outer south west)--(myPlots c5r2.outer south east)
		node {Transformation Coefficients}
		;
		
		\end{tikzpicture}
	}
	
	\caption{Independent and uniform spatial scaling and shearing of the X, Y and Z axes. Top row plots illustrate experimentally captured results. Bottom row plots illustrate transformation estimations using Equation \ref{eq:dft_3D_transformed}, with discrete integer estimation illustrated by a solid line and continuous floating-point estimation by a dotted line.}
	\label{plots:granularScaleShearPlots}
\end{figure}
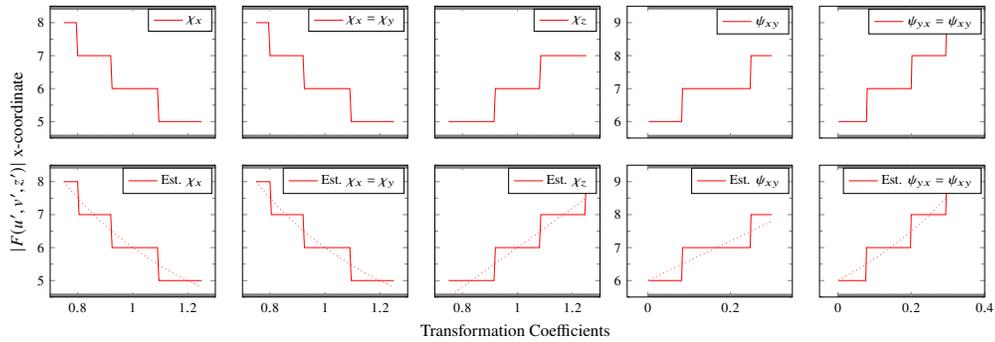

To demonstrate continuity in conformance, samples were collected for a granular range of coefficients, sampling independent and uniform rotation, scaling $\chi_x$, $\chi_y$, $\chi_z$, $\chi_x\!=\!\chi_y$, and $\chi_x\!=\!\chi_y\!=\!\chi_z$, and shearing $\psi_{yx}$, $\psi_{xy}$, and $\psi_{yx}\!=\!\psi_{xy}$. Using the experimental platform previously described, scaling coefficients were incremented by 0.005 between 0.75 and 1.25 for a total of 101 samples, shearing coefficients were incremented by 0.002 between 0.002 and 0.3 for a total of 150 samples, and rotation was sampled at 1 degree increments between 0 and 360 degrees. A subset of these tables are plotted in Figure \ref{plots:granularScaleShearPlots}, showing the congruence between captured and estimated points in $|F(u',v',z')|$ images for a range of transformation coefficients.

\section{Special Cases}
\label{sec:LimitationAndSpecialCases}

Special cases exist in Theorem~\ref{PTtheorem} for SFTP which include spatial translations and transformations in higher dimensional space. Such spatial transformations alter the representation of the frequency domain, and visualization of the spectrum, as previous done using the magnitude of Equation~\ref{eq:dft_3D_transformed_Mag}, may not allow for a direct inference of the transformation. This section outlines transformation inference for these special cases, using Perspective Transform Theorem for SFTP as expressed in Equation~\ref{eq:dft_3D_transformed}. 

\subsection{Translation}
\label{subsec:translation}
Spatial translation does not manifest as a coordinate system transformation in the Fourier domain. Though expressed mathematically in Equation~\ref{eq:dft_3D_transformed}, spatial translation coefficients are omitted from the coordinate system transformation defined in Equation~\ref{eq:dft_3D_transformed_Mag}. This is a result of spatial translation representing a complex and spatially aware scaling of $F(u,v,w)$ which can be isolated and expressed through Equation~\ref{eq:dft_3D_translation}:

\begin{equation}
G(u,v, w) = F(u,v,w)e^{2\pi\imath(EC)}
\label{eq:dft_3D_translation}
\end{equation}

In DFTs, the impact of spatial transformations on the Fourier domain is proportional to the period of the sampled region. Defined by the relationships in Equation~\ref{eq:periodicity_spatial} and \ref{eq:periodicity_fourier}, translation along any dimension in multiples of the dimension's period will not result in alteration of the Fourier representation~\cite[~p.210]{book:digimproc2ed}. Translation in any dimension by increments other than the dimension's period will result in a quantifiable shift in the phase of Fourier domain frequencies.

\begin{equation} 
f(x,y) = 
f(x+M,y) = f(x,y+N) = f(x+M,y+N)
\label{eq:periodicity_spatial}
\end{equation}

\begin{equation} 
F(u,v) = 
F(u+M,v) = F(u,v+N) = F(u+M,v+N)
\label{eq:periodicity_fourier}
\end{equation}

To account for periodicity in DFTs, translation coefficients are scaled by the length of their respective dimensions $M$, $N$, and $D$ as previously defined in Equation~\ref{eq:dft_spatial_to_frequency} and applied in $E$ of Equation~\ref{eq:dft_3D_translation}. This scaling is not applied to $H$ in Equation~\ref{eq:dft_3D_transformed_Mag} as the input points originate in Fourier space, and phase shifts are invariant in the magnitude of a DFT.

\subsection{Validation of Translation}

\begin{figure}[!htb]
	\centering
	\subfloat[SB]{
		\includegraphics[width=0.25\textwidth]{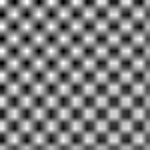}
		\label{fig:Translate_XY_by6_rawSB}
	}
	\subfloat[$Re(\mathcal{F}(SB))$]{
		\includegraphics[width=0.25\textwidth]{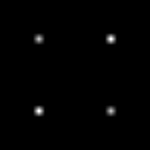}
		\label{fig:Translate_XY_by6_Re(F(SB))}
	}
	\subfloat[$Im(\mathcal{F}(SB))$]{
		\includegraphics[width=0.25\textwidth]{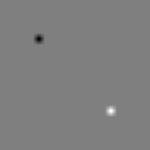}
		\label{fig:Translate_XY_by6_Im(F(SB))}
	}
	\subfloat[$|\mathcal{F}(SB)|$]{
		\includegraphics[width=0.25\textwidth]{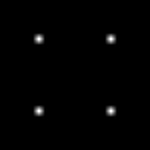}
		\label{fig:Translate_XY_by6_Mag(F(SB))}
	}	
	\caption{Uniform spatial translation by six pixels $\tau_{x}\!=\!\tau_{y}=6$}
	\label{fig:Translate_XY_by6}
\end{figure}

To validate Equation~\ref{eq:dft_3D_translation} for translation, $25\!\times\!25$ spatial samples were collected from $75\!\times\!75$ images which were encoded in the frequency domain. Applying Equation~\ref{eq:periodicity_fourier} for periodicity, encoding was implemented such that $25\!\times\!25$ samples of the image consisted of four frequencies with amplitude 10,000 at Cartesian coordinates $(6,6)$,$(-6,-6)$,$(6,-6)$,
and $(-6,6)$ of the Fourier spectrum. This periodicity allowed for mitigation of noise from pattern truncation and for sampling of the central $25\!\times\!25$ region of the encoded image at one pixel increments from 0 to 25 pixels in the same manner as described in Section~\ref{sec:experimentalResults}. Figure~\ref{fig:Translate_XY_by6} illustrates a sample from this set. 

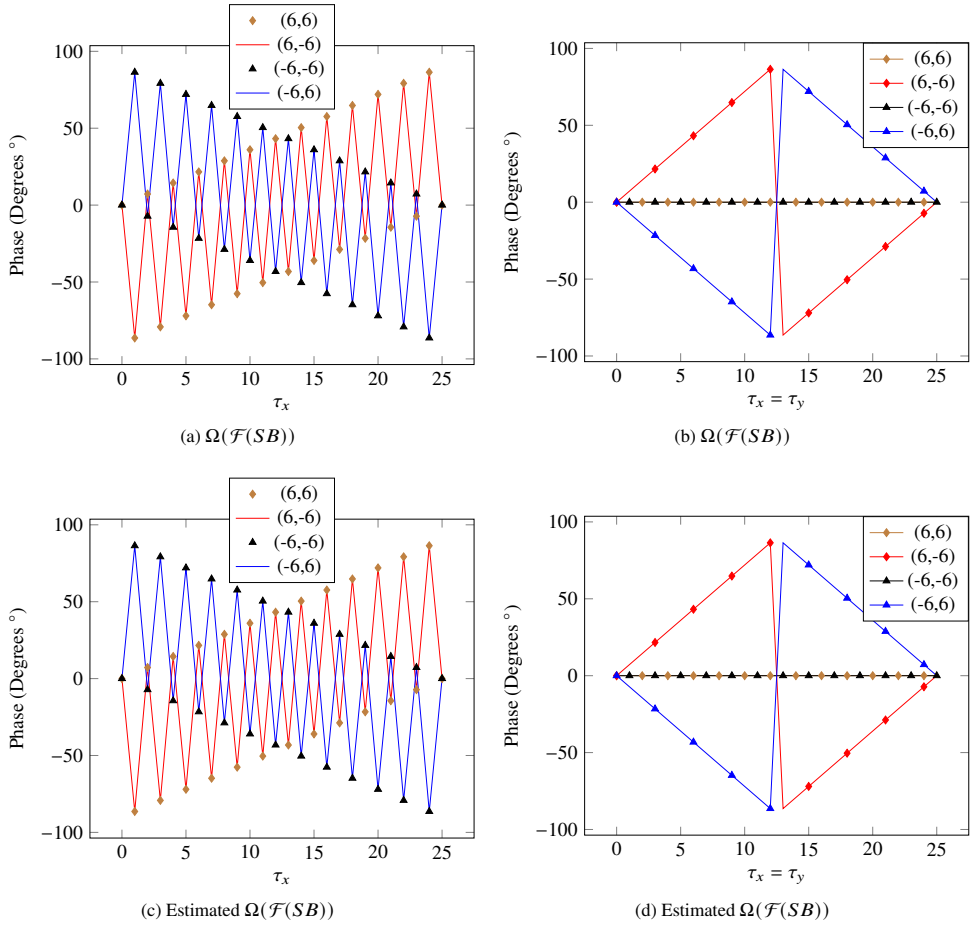
\begin{figure}[!ht]
	\centering
	\subfloat[$\Omega(\mathcal{F}(SB))$]{
		\resizebox {0.48\textwidth} {!} {
			\begin{tikzpicture}
			\begin{axis}[
			ylabel={Phase (Degrees $^\circ$)},
			xlabel={$\tau_{x}$},
			legend style={at={(0.5,0.8)},anchor=south}]
			\addplot [style={solid}, only marks, mark=diamond*, color=brown] table [col sep=comma, x=Translate_Tx, y={Filt_PhaseCalc[6 6]}]{Translate_Tx.csv};
			\addlegendentry{(6,6)}
			
			\addplot [style={solid}, mark=none, color=red] table [col sep=comma, x=Translate_Tx, y={Filt_PhaseCalc[6 -6]}]{Translate_Tx.csv};
			\addlegendentry{(6,-6)}
			
			\addplot [style={solid}, only marks, mark=triangle*, color=black] table [col sep=comma, x=Translate_Tx, y={Filt_PhaseCalc[-6 -6]}]{Translate_Tx.csv};
			\addlegendentry{(-6,-6)}
			
			\addplot [style={solid}, mark=none, color=blue] table [col sep=comma, x=Translate_Tx, y={Filt_PhaseCalc[-6 6]}]{Translate_Tx.csv};
			\addlegendentry{(-6,6)}
			\end{axis}
			\end{tikzpicture}
			\label{plot:SB_TranslationX_Phase}
		}
	}
	\subfloat[$\Omega(\mathcal{F}(SB))$]{
		\resizebox {0.48\textwidth} {!} {
			\begin{tikzpicture}
			\begin{axis}[
			ylabel={Phase (Degrees $^\circ$)},
			xlabel={$\tau_{x}=\tau_{y}$},
			legend style={at={(1,1)},anchor={north east}}]
			\addplot [style={solid}, mark=diamond*, color=brown, mark repeat=2,mark phase=0] table [col sep=comma, x=Translate_TxTy, y={Filt_PhaseCalc[6 6]}]{Translate_TxTy.csv};
			\addlegendentry{(6,6)}
			\addplot [style={solid}, mark=diamond*, color=red, mark repeat=3, mark phase=1] table [col sep=comma, x=Translate_TxTy, y={Filt_PhaseCalc[6 -6]}]{Translate_TxTy.csv};
			\addlegendentry{(6,-6)}
			\addplot [style={solid}, mark=triangle*, color=black, mark repeat=2,mark phase=2] table [col sep=comma, x=Translate_TxTy, y={Filt_PhaseCalc[-6 -6]}]{Translate_TxTy.csv};
			\addlegendentry{(-6,-6)}
			\addplot [style={solid}, mark=triangle*, color=blue, mark repeat=3, mark phase=1] table [col sep=comma, x=Translate_TxTy, y={Filt_PhaseCalc[-6 6]}]{Translate_TxTy.csv};
			\addlegendentry{(-6,6)}
			\end{axis}
			\end{tikzpicture}
			\label{plot:SB_TranslationXY_Phase}
		}
	}
	\hfill
	\subfloat[Estimated $\Omega(\mathcal{F}(SB))$]{
		\resizebox {0.48\textwidth} {!} {
			\begin{tikzpicture}
			\begin{axis}[
			ylabel={Phase (Degrees $^\circ$)},
			xlabel={$\tau_{x}$},
			legend style={at={(0.5,0.8)},anchor=south}]
			\addplot [style={solid}, only marks, mark=diamond*, color=brown] table [col sep=comma, x=Translate_Tx, y={Est_EstPhaseCalc[6 6]}]{Translate_Tx.csv};
			\addlegendentry{(6,6)}
			
			\addplot [style={solid}, mark=none, color=red] table [col sep=comma, x=Translate_Tx, y={Est_EstPhaseCalc[6 -6]}]{Translate_Tx.csv};
			\addlegendentry{(6,-6)}
			
			\addplot [style={solid}, only marks, mark=triangle*, color=black] table [col sep=comma, x=Translate_Tx, y={Est_EstPhaseCalc[-6 -6]}]{Translate_Tx.csv};
			\addlegendentry{(-6,-6)}
			
			\addplot [style={solid}, mark=none, color=blue] table [col sep=comma, x=Translate_Tx, y={Est_EstPhaseCalc[-6 6]}]{Translate_Tx.csv};
			\addlegendentry{(-6,6)}
			\end{axis}
			\end{tikzpicture}
			\label{plot:SB_TranslationX_PhaseEst}
		}
	}
	\subfloat[Estimated $\Omega(\mathcal{F}(SB))$]{
		\resizebox {0.48\textwidth} {!} {
			\begin{tikzpicture}
			\begin{axis}[
			ylabel={Phase (Degrees $^\circ$)},
			xlabel={$\tau_{x}=\tau_{y}$},
			legend style={at={(1,1)},anchor={north east}}]
			\addplot [style={solid}, mark=diamond*, color=brown, mark repeat=2,mark phase=0] table [col sep=comma, x=Translate_TxTy, y={Est_EstPhaseCalc[6 6]}]{Translate_TxTy.csv};
			\addlegendentry{(6,6)}
			\addplot [style={solid}, mark=diamond*, color=red, mark repeat=3, mark phase=1] table [col sep=comma, x=Translate_TxTy, y={Est_EstPhaseCalc[6 -6]}]{Translate_TxTy.csv};
			\addlegendentry{(6,-6)}
			\addplot [style={solid}, mark=triangle*, color=black, mark repeat=2,mark phase=2] table [col sep=comma, x=Translate_TxTy, y={Est_EstPhaseCalc[-6 -6]}]{Translate_TxTy.csv};
			\addlegendentry{(-6,-6)}
			\addplot [style={solid}, mark=triangle*, color=blue, mark repeat=3, mark phase=1] table [col sep=comma, x=Translate_TxTy, y={Est_EstPhaseCalc[-6 6]}]{Translate_TxTy.csv};
			\addlegendentry{(-6,6)}
			\end{axis}
			\end{tikzpicture}
			\label{plot:SB_TranslationXY_PhaseEst}
		}
	}
	\caption{Phase angles of encoded points from $25\times 25$ samples translated by 0 to 25 pixels.}
	\label{plot:translation_Phase}
\end{figure}

The impact of Equation~\ref{eq:dft_3D_translation} on the Fourier spectrum can be measured by calculating the phase angles of each encoded point using Equation~\ref{eq:ft_phase}. To validate, Figure~\ref{plot:translation_Phase} illustrates the impact of spatial translation on the $x$-axis ($\tau_{x}$) and uniformly on the $x$ and $y$-axes ($\tau_{x}\!=\!\tau_{y}$) for one period of the encoded pattern. The magnitude of the spectrum remains constant throughout, and conformity between experimental and mathematically estimated results is illustrated respectively by Figures \ref{plot:SB_TranslationX_Phase} and \ref{plot:SB_TranslationX_PhaseEst} and Figures \ref{plot:SB_TranslationXY_Phase} and \ref{plot:SB_TranslationXY_PhaseEst}.

\begin{equation}
\Omega(u,v,w) = \arctan\left(\frac{I(u,v,w)}{R(u,v,w)}\right) 
\label{eq:ft_phase}
\end{equation}

The Fourier spectrum has a $0.5\!\times\!(M\!\times\!N)$ periodicity for $M\!\times\!N$ images and thus reflects across the zero frequency (Cartesian origin). Encoded points in the Cartesian quadrants I and III are the product of reflection and can be subject to destructive interference as seen in Figure~\ref{fig:Translate_XY_by6_Im(F(SB))} and the plots in Figure~\ref{plot:translation_Phase}. Conformity between phase angles calculated through Equation~\ref{eq:dft_3D_translation} and experimental results in Figure~\ref{plot:translation_Phase} illustrate how translation is a measure of periodic misalignment and further validates the Perspective Transformation Theorem for translation SFTP.

\subsection{Shearing Onto the Z-axis}
\label{subsec:warp}

In 2D space, transformations such as a shearing of the $x$ and $y$-axes onto the $z$-axis (warp, $\psi_{xz}$ and $\psi_{yz}$) transform 2D input in 3D space. Often as an implicit step, a reduction in dimensionality is required to reduce the 3D result to the 2D dimensionality of the original input. The approach to validation used in this work simulates the projection and subsequent image capture of 2D patterns, allowing for projection in 3D space with implicit reduction to 2D space on re-capture. This implicit operation is synonymous with the application of Equation~\ref{eq:perspective_divide} after transformation, which can otherwise be restated for an image as Equation~\ref{eq:3D_to_2D_reduction}.

\begin{equation}
f\left(\frac{x'}{z'},\frac{y'}{z'},1\right) = f\left(\frac{x'}{z'},\frac{y'}{z'},\frac{z'}{z'}\right) = f(x',y',z') 
\label{eq:3D_to_2D_reduction}
\end{equation}

As a result, a 2D spatial image warped by $f(x,y,({\psi}_{xz}x + {\psi}_{yz}y + z))$ is implicitly transformed by Equation \ref{eq:3D_to_2D_reduction} if 2D dimensionality is to be retained. This effectively alters the warp transformation to reflect a uniform scaling of the $x$ and $y$-axes as shown in Equation \ref{eq:2Dspatial_warp}. Inherent to this transformation, an interpolation algorithm is required to resolve the many-to-one pixel mappings from $f(x,y,z)$ to $f(x',y',z')$, for which many algorithms exist with varied pixel-binning characteristics. 

\begin{equation}
f(x',y',1) = f\left(\frac{x}{({\psi}_{xz}x + {\psi}_{yz}y + z)},\frac{y}{({\psi}_{xz}x + {\psi}_{yz}y + z)}, \frac{({\psi}_{xz}x + {\psi}_{yz}y + z)}{({\psi}_{xz}x + {\psi}_{yz}y + z)}\right)
\label{eq:2Dspatial_warp}
\end{equation}

\begin{figure}[!ht]
	\centering
	\subfloat[SB]{
		\includegraphics[width=0.2\textwidth]{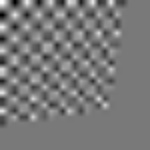}
		\label{fig:BLI_Warp_XY_by0_0075_rawSB}
	}
	\subfloat[$Re(\mathcal{F}(SB))$]{
		\includegraphics[width=0.2\textwidth]{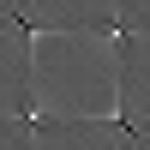}
		\label{fig:BLI_Warp_XY_by0_0075_rawF_SB__Real}
	}
	\subfloat[$Im(\mathcal{F}(SB))$]{
		\includegraphics[width=0.2\textwidth]{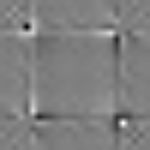}
		\label{fig:BLI_Warp_XY_by0_0075_rawF_SB__Imaginary}
	}
	\subfloat[$|\mathcal{F}(SB)|$]{
		\includegraphics[width=0.2\textwidth]{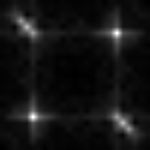}
		\label{fig:BLI_Warp_XY_by0_0075_rawF(SB)}
	}
	\hfill
	\subfloat[SB]{
		\includegraphics[width=0.2\textwidth]{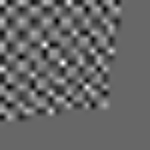}
		\label{fig:Warp_XY_by0_0075_rawSB}
	}
	\subfloat[$Re(\mathcal{F}(SB))$]{
		\includegraphics[width=0.2\textwidth]{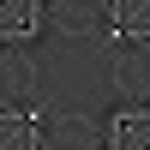}
		\label{fig:Warp_XY_by0_0075_rawF_SB__Real}
	}
	\subfloat[$Im(\mathcal{F}(SB))$]{
		\includegraphics[width=0.2\textwidth]{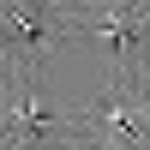}
		\label{fig:Warp_XY_by0_0075_rawF_SB__Imaginary}
	}
	\subfloat[$|\mathcal{F}(SB)|$]{
		\includegraphics[width=0.2\textwidth]{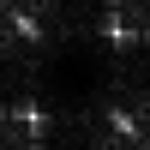}
		\label{fig:Warp_XY_by0_0075_rawF(SB)}
	}
	\caption{A uniform spatial warp of the X \&Y axes by 0.0075 ($\psi_{xz}=\psi_{yz}=0.0075$). Figures \ref{fig:BLI_Warp_XY_by0_0075_rawSB}-\ref{fig:BLI_Warp_XY_by0_0075_rawF_SB__Real} use bilinear interpolation and Figures \ref{fig:Warp_XY_by0_0075_rawSB}-\ref{fig:Warp_XY_by0_0075_rawF_SB__Real} were interpolated using an arithmetic sum.}
	\label{fig:Warp_XY}
\end{figure}

In the Fourier spectrum, this spatial transformation produces the complex conjugate shown in Equation~\ref{eq:dft_warp} which preserves points $F(u,v,w)$ while introducing spatially-aware scaling as a factor of the $x$ and $y$-axes. This phenomenon is illustrated in Figures~\ref{fig:Warp_XY} of a uniform spatial warp by 0.0075 ($\psi_{xz}=\psi_{yz}=0.0075$). Figures~\ref{fig:Warp_XY_by0_0075_rawF_SB__Real}-\ref{fig:Warp_XY_by0_0075_rawF(SB)} illustrate the Fourier representation of the spatial image, including the smearing of encoded points towards higher frequencies in $F(u,v)$ space. This scaling is prominent in the II and IV Cartesian quadrants, though horizontal and vertical smearing can be seen in Figure~\ref{fig:Warp_XY_by0_0075_rawF(SB)} for all encoded points.

\begin{gather}
F(u',v',w') = JF(u,v,w) \label{eq:dft_warp}\\
\text{where }J=
\begin{bmatrix}
{u} && {v} && {w}
\end{bmatrix}
\begin{bmatrix}
\frac{1}{({\psi}_{xz}x + {\psi}_{yz}y + z)} 	&& 0	&& 	0	\\
0	&& \frac{1}{({\psi}_{xz}x + {\psi}_{yz}y + z)}	&& 	0	\\
0	&& 0	&&	\frac{({\psi}_{xz}x + {\psi}_{yz}y + z)}{({\psi}_{xz}x + {\psi}_{yz}y + z)}
\end{bmatrix}^{-1}
\label{eq:dft_warp_J}
\end{gather}

\subsection{Validation of Warp}

To validate SFTP perspective theorem for spatial warp transformations, $25\!\times\!25$ spatial images were encoded in the frequency domain with four frequencies of amplitude 10,000 at Cartesian coordinates $(6,6)$, $(-6,-6)$, $(6,-6)$ and $(-6,6)$ of the Fourier spectrum $F(u,v,w)$ as previously described in Section~\ref{sec:experimentalResults}. Figure~\ref{fig:Warp_XY} illustrates a bilinearly interpolated and summing interpolation samples from this set, including the transformed spatial images in Figure~\ref{fig:BLI_Warp_XY_by0_0075_rawSB} and \ref{fig:Warp_XY_by0_0075_rawSB}. The spatial representation shows the image deformation and the introduction of blank space (constant grey) to the right and bottom which will be represented as noise in frequency domain. This blank space cannot be mitigated through periodicity due to the spatially-aware nature of spatial warp transformations (see Equation~\ref{eq:dft_warp}). 

\begin{table}[!ht]
	\centering
	\scriptsize
	\caption{Spatial shearing onto the Z-Axis (Warp)}
	\begin{tabular}{c|c|c|c}
		\hline
		\textbf{Spatial Transform} &  \textbf{$|\mathcal{F}(SB)|$ Captured} & \textbf{$|\mathcal{F}(SB)|$ Calculated} & \textbf{Estimated Transform}
		\\ \hline
		$None$
		& \begin{tabular}[c]{@{}c@{}}(6, 6)(-6, -6)\\(6, -6)(-6, 6)\end{tabular}
		& \begin{tabular}[c]{@{}c@{}}(6, 6)(-6, -6)\\(6, -6)(-6, 6)\end{tabular}
		& $None$	
		\\ \hline
		${\psi}_{zx}=0.003$	
		& \begin{tabular}[c]{@{}c@{}}(6, 6)(-6,-6)\\(6, -6)(-6, 6)\end{tabular}
		& \begin{tabular}[c]{@{}c@{}}(6.003, 6)(-5.997, -6)\\(6.003, -6)(-5.997, 6)\end{tabular}
		& ${\psi}_{zx}=0.003$	
		\\ \hline
		${\psi}_{zy}=0.003$
		& \begin{tabular}[c]{@{}c@{}}(6, 6)(-6,-6)\\(6, -6)(-6, 6)\end{tabular}
		& \begin{tabular}[c]{@{}c@{}}(6, 5.997)(-6, -6.003)\\(6, -6.003)(-6, 5.997)\end{tabular}
		& ${\psi}_{zy}=0.003$	
		\\ \hline
		${\psi}_{zx}={\psi}_{zy}=0.003$	
		& \begin{tabular}[c]{@{}c@{}}(6, 6)(-6, -6)\\(7, -7)(-7, 7)\end{tabular}
		& \begin{tabular}[c]{@{}c@{}}(6.003, 5.997)(-5.997, -6.003)\\(6.003, -6.003)(-5.997, 5.997)\end{tabular}
		& ${\psi}_{zx}={\psi}_{zy}=0.003$	
		\\ \hline
	\end{tabular}
	\label{table:Warp}
\end{table}

\begin{figure}[!ht]
	\centering
	\subfloat[$\Omega(\mathcal{F}(SB))$]{
		\resizebox {0.35\textwidth} {!} {
			\begin{tikzpicture}
			\begin{axis}[
			ylabel={Phase (Degrees $^\circ$)},
			xlabel={$\psi_{xz}$},
			legend style={at={(1,1)},anchor={north east}}]
			\addplot [style={dotted}, mark=none, color=red, forget plot]  table [col sep=comma, x=Warp_X, y={Initial_PhaseCalc[6 6]}]{Warp_X.csv};
			\addplot [style={dotted}, mark=none, color=blue, forget plot] table [col sep=comma, x=Warp_X, y={Initial_PhaseCalc[-6 -6]}]{Warp_X.csv};
			\addplot [style={dotted}, mark=none, color=brown, forget plot]  table [col sep=comma, x=Warp_X, y={Initial_PhaseCalc[6 -6]}]{Warp_X.csv};
			\addplot [style={dotted}, mark=none, color=green, forget plot] table [col sep=comma, x=Warp_X, y={Initial_PhaseCalc[-6 6]}]{Warp_X.csv};
			
			\addplot [style={solid}, mark=none, color=red] table [col sep=comma, x=Warp_X, y={Mag_PhaseCalc[6 6]}]{Warp_X.csv};
			\addlegendentry{(6,6)}
			\addplot [style={solid}, mark=none, color=blue] table [col sep=comma, x=Warp_X, y={Mag_PhaseCalc[-6 -6]}]{Warp_X.csv};
			\addlegendentry{(-6,-6)}
			\addplot [style={solid}, mark=none, color=brown] table [col sep=comma, x=Warp_X, y={Mag_PhaseCalc[-6 6]}]{Warp_X.csv};
			\addlegendentry{(-6,6)}
			\addplot [style={solid}, mark=none, color=green] table [col sep=comma, x=Warp_X, y={Mag_PhaseCalc[6 -6]}]{Warp_X.csv};
			\addlegendentry{(6,-6)}
			\end{axis}
			\end{tikzpicture}
			\label{plot:WarpX_Phase}
		}
	}
	\subfloat[$|\mathcal{F}(SB)|$]{
		\resizebox {0.35\textwidth} {!} {
			\begin{tikzpicture}
			\begin{axis}[
			ylabel={Magnitude},
			xlabel={$\psi_{xz}$},
			legend style={at={(1,1)},anchor={north east}}]
			\addplot [style={dotted}, mark=none, color=red, forget plot] table [col sep=comma, x=Warp_X, y={Initial_MagVal[6 6]}]{Warp_X.csv};
			\addplot [style={dotted}, mark=none, color=blue, forget plot] table [col sep=comma, x=Warp_X, y={Initial_MagVal[-6 -6]}]{Warp_X.csv};
			\addplot [style={dotted}, mark=none, color=brown, forget plot]  table [col sep=comma, x=Warp_X, y={Initial_MagVal[6 -6]}]{Warp_X.csv};
			\addplot [style={dotted}, mark=none, color=green, forget plot] table [col sep=comma, x=Warp_X, y={Initial_MagVal[-6 6]}]{Warp_X.csv};
			
			\addplot [style={solid}, mark=none, color=red] table [col sep=comma, x=Warp_X, y={Mag_MagVal[6 6]}]{Warp_X.csv};
			\addlegendentry{(6,6)}
			\addplot [style={solid}, mark=none, color=blue] table [col sep=comma, x=Warp_X, y={Mag_MagVal[-6 -6]}]{Warp_X.csv};
			\addlegendentry{(-6,-6)}
			\addplot [style={solid}, mark=none, color=brown] table [col sep=comma, x=Warp_X, y={Mag_MagVal[6 -6]}]{Warp_X.csv};
			\addlegendentry{(6,-6)}
			\addplot [style={solid}, mark=none, color=green] table [col sep=comma, x=Warp_X, y={Mag_MagVal[-6 6]}]{Warp_X.csv};
			\addlegendentry{(-6,6)}
			\end{axis}
			\end{tikzpicture}
			\label{plot:WarpX_Magnitude}
		}
	}
	\hfill
	\subfloat[Estimated $\Omega(\mathcal{F}(SB))$]{
		\resizebox {0.35\textwidth} {!} {
			\begin{tikzpicture}
			\begin{axis}[
			ylabel={Phase (Degrees $^\circ$)},
			xlabel={$\psi_{xz}$},
			legend style={at={(1,1)},anchor={north east}}]
			
			\addplot [style={solid}, mark=none, color=red] table [col sep=comma, x=Warp_X, y={Est_EstPhaseCalc[6 6]}]{Warp_X.csv};
			\addlegendentry{(6,6)}
			\addplot [style={solid}, mark=none, color=blue] table [col sep=comma, x=Warp_X, y={Est_EstPhaseCalc[-6 -6]}]{Warp_X.csv};
			\addlegendentry{(-6,-6)}
			\addplot [style={solid}, mark=none, color=brown] table [col sep=comma, x=Warp_X, y={Est_EstPhaseCalc[-6 6]}]{Warp_X.csv};
			\addlegendentry{(-6,6)}
			\addplot [style={solid}, mark=none, color=green] table [col sep=comma, x=Warp_X, y={Est_EstPhaseCalc[6 -6]}]{Warp_X.csv};
			\addlegendentry{(6,-6)}
			\end{axis}
			\end{tikzpicture}
			\label{plot:EstWarpX_Phase}
		}
	}
	\subfloat[Estimated $|\mathcal{F}(SB)|$]{
		\resizebox {0.35\textwidth} {!} {
			\begin{tikzpicture}
			\begin{axis}[
			ylabel={Magnitude},
			xlabel={$\psi_{xz}$},
			legend style={at={(1,1)},anchor={north east}}]
			
			\addplot [style={solid}, mark=none, color=red] table [col sep=comma, x=Warp_X, y={Est_EstMagCalc[6 6]}]{Warp_X.csv};
			\addlegendentry{(6,6)}
			\addplot [style={solid}, mark=none, color=blue] table [col sep=comma, x=Warp_X, y={Est_EstMagCalc[-6 -6]}]{Warp_X.csv};
			\addlegendentry{(-6,-6)}
			\addplot [style={solid}, mark=none, color=brown] table [col sep=comma, x=Warp_X, y={Est_EstMagCalc[6 -6]}]{Warp_X.csv};
			\addlegendentry{(6,-6)}
			\addplot [style={solid}, mark=none, color=green] table [col sep=comma, x=Warp_X, y={Est_EstMagCalc[-6 6]}]{Warp_X.csv};
			\addlegendentry{(-6,6)}
			\end{axis}
			\end{tikzpicture}
			\label{plot:EstWarpX_Magnitude}
		}
	}
    \hfill
	\subfloat[$\Omega(\mathcal{F}(SB))$]{
		\resizebox {0.35\textwidth} {!} {
			\begin{tikzpicture}
			\begin{axis}[
			ylabel={Phase (Degrees $^\circ$)},
			xlabel={$\psi_{xz}=\psi_{yz}$},
			legend style={at={(1,1)},anchor={north east}}]
			\addplot [style={dotted}, mark=none, color=red, forget plot]  table [col sep=comma, x=Warp_XY, y={Initial_PhaseCalc[6 6]}]{Warp_XY.csv};
			\addplot [style={dotted}, mark=none, color=blue, forget plot] table [col sep=comma, x=Warp_XY, y={Initial_PhaseCalc[-6 -6]}]{Warp_XY.csv};
			\addplot [style={dotted}, mark=none, color=brown, forget plot]  table [col sep=comma, x=Warp_XY, y={Initial_PhaseCalc[6 -6]}]{Warp_XY.csv};
			\addplot [style={dotted}, mark=none, color=green, forget plot] table [col sep=comma, x=Warp_XY, y={Initial_PhaseCalc[-6 6]}]{Warp_XY.csv};
			
			\addplot [style={solid}, mark=none, color=red] table [col sep=comma, x=Warp_XY, y={Mag_PhaseCalc[6 6]}]{Warp_XY.csv};
			\addlegendentry{(6,6)}
			\addplot [style={solid}, mark=none, color=blue] table [col sep=comma, x=Warp_XY, y={Mag_PhaseCalc[-6 -6]}]{Warp_XY.csv};
			\addlegendentry{(-6,-6)}
			\addplot [style={solid}, mark=none, color=brown] table [col sep=comma, x=Warp_XY, y={Mag_PhaseCalc[-6 6]}]{Warp_XY.csv};
			\addlegendentry{(-6,6)}
			\addplot [style={solid}, mark=none, color=green] table [col sep=comma, x=Warp_XY, y={Mag_PhaseCalc[6 -6]}]{Warp_XY.csv};
			\addlegendentry{(6,-6)}
			\end{axis}
			\end{tikzpicture}
			\label{plot:WarpXY_Phase}
		}
	}
	\subfloat[$|\mathcal{F}(SB)|$]{
		\resizebox {0.35\textwidth} {!} {
			\begin{tikzpicture}
			\begin{axis}[
			ylabel={Magnitude},
			xlabel={$\psi_{xz}=\psi_{yz}$},
			legend style={at={(1,1)},anchor={north east}}]
			\addplot [style={dotted}, mark=none, color=red, forget plot] table [col sep=comma, x=Warp_XY, y={Initial_MagVal[6 6]}]{Warp_XY.csv};
			\addplot [style={dotted}, mark=none, color=blue, forget plot] table [col sep=comma, x=Warp_XY, y={Initial_MagVal[-6 -6]}]{Warp_XY.csv};
			\addplot [style={dotted}, mark=none, color=brown, forget plot]  table [col sep=comma, x=Warp_XY, y={Initial_MagVal[6 -6]}]{Warp_XY.csv};
			\addplot [style={dotted}, mark=none, color=green, forget plot] table [col sep=comma, x=Warp_XY, y={Initial_MagVal[-6 6]}]{Warp_XY.csv};
			
			\addplot [style={solid}, mark=none, color=red] table [col sep=comma, x=Warp_XY, y={Mag_MagVal[6 6]}]{Warp_XY.csv};
			\addlegendentry{(6,6)}
			\addplot [style={solid}, mark=none, color=blue] table [col sep=comma, x=Warp_XY, y={Mag_MagVal[-6 -6]}]{Warp_XY.csv};
			\addlegendentry{(-6,-6)}
			\addplot [style={solid}, mark=none, color=brown] table [col sep=comma, x=Warp_XY, y={Mag_MagVal[6 -6]}]{Warp_XY.csv};
			\addlegendentry{(6,-6)}
			\addplot [style={solid}, mark=none, color=green] table [col sep=comma, x=Warp_XY, y={Mag_MagVal[-6 6]}]{Warp_XY.csv};
			\addlegendentry{(-6,6)}
			\end{axis}
			\end{tikzpicture}
			\label{plot:WarpXY_Magnitude}
		}
	}
	\hfill
	\subfloat[Estimated $\Omega(\mathcal{F}(SB))$]{
		\resizebox {0.35\textwidth} {!} {
			\begin{tikzpicture}
			\begin{axis}[
			ylabel={Phase (Degrees $^\circ$)},
			xlabel={$\psi_{xz}=\psi_{yz}$},
			legend style={at={(1,1)},anchor={north east}}]
			
			\addplot [style={solid}, mark=none, color=red] table [col sep=comma, x=Warp_XY, y={Est_EstPhaseCalc[6 6]}]{Warp_XY.csv};
			\addlegendentry{(6,6)}
			\addplot [style={solid}, mark=none, color=blue] table [col sep=comma, x=Warp_XY, y={Est_EstPhaseCalc[-6 -6]}]{Warp_XY.csv};
			\addlegendentry{(-6,-6)}
			\addplot [style={solid}, mark=none, color=brown] table [col sep=comma, x=Warp_XY, y={Est_EstPhaseCalc[-6 6]}]{Warp_XY.csv};
			\addlegendentry{(-6,6)}
			\addplot [style={solid}, mark=none, color=green] table [col sep=comma, x=Warp_XY, y={Est_EstPhaseCalc[6 -6]}]{Warp_XY.csv};
			\addlegendentry{(6,-6)}
			\end{axis}
			\end{tikzpicture}
			\label{plot:EstWarpXY_Phase}
		}
	}
	\subfloat[Estimated $|\mathcal{F}(SB)|$]{
		\resizebox {0.35\textwidth} {!} {
			\begin{tikzpicture}
			\begin{axis}[
			ylabel={Magnitude},
			xlabel={$\psi_{xz}=\psi_{yz}$},
			legend style={at={(1,1)},anchor={north east}}]
			
			\addplot [style={solid}, mark=none, color=red] table [col sep=comma, x=Warp_XY, y={Est_EstMagCalc[6 6]}]{Warp_XY.csv};
			\addlegendentry{(6,6)}
			\addplot [style={solid}, mark=none, color=blue] table [col sep=comma, x=Warp_XY, y={Est_EstMagCalc[-6 -6]}]{Warp_XY.csv};
			\addlegendentry{(-6,-6)}
			\addplot [style={solid}, mark=none, color=brown] table [col sep=comma, x=Warp_XY, y={Est_EstMagCalc[6 -6]}]{Warp_XY.csv};
			\addlegendentry{(6,-6)}
			\addplot [style={solid}, mark=none, color=green] table [col sep=comma, x=Warp_XY, y={Est_EstMagCalc[-6 6]}]{Warp_XY.csv};
			\addlegendentry{(-6,6)}
			\end{axis}
			\end{tikzpicture}
			\label{plot:EstWarpXY_Magnitude}
		}
	}
	\caption{Spatial shearing of the X and Y-axes onto the Z-axis (warp, $\psi_{xz}$ and $\psi_{xz}=\psi_{yz}$) sampled at increments of 0.00005 from 0.0 to 0.01}
	\label{plots:WarpXY}
\end{figure}
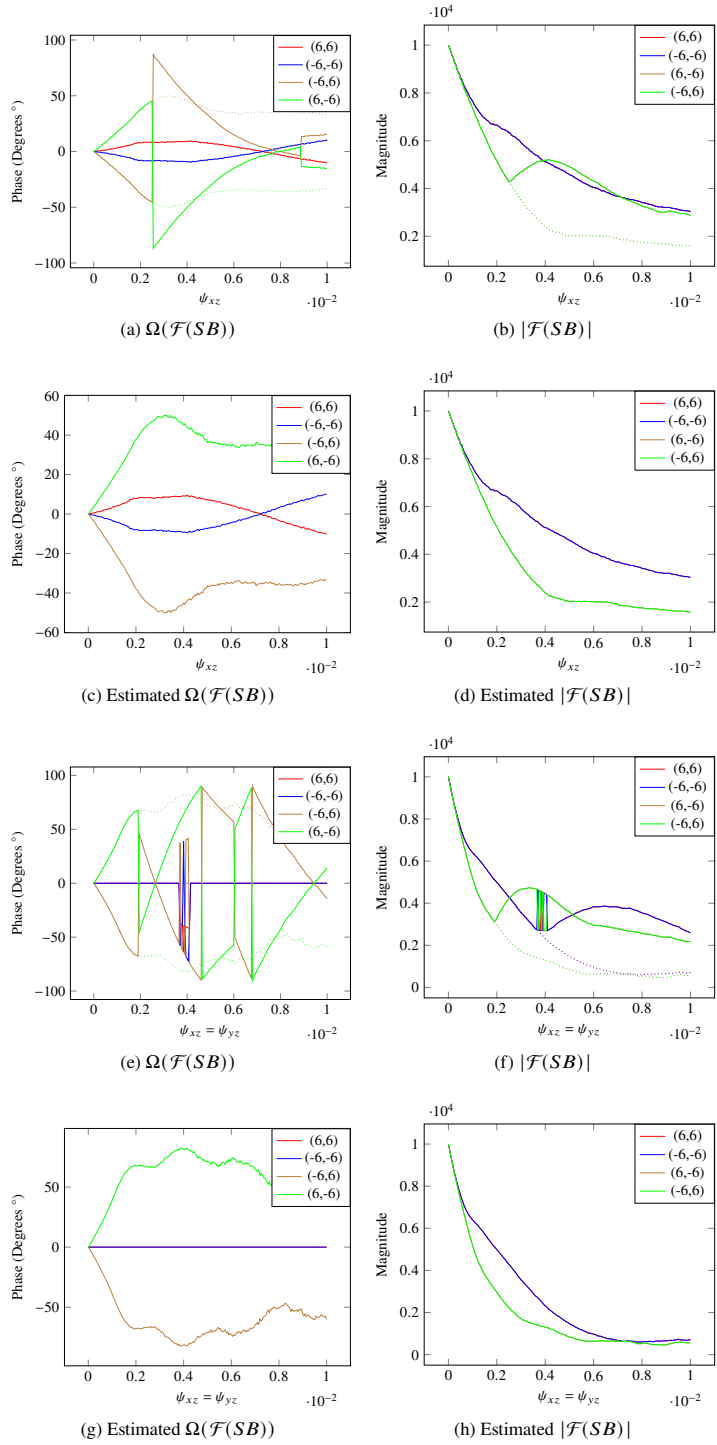

To validate uniform scaling of the frequency domain by spatial warp and ascertain conformity with Equation~\ref{eq:dft_warp}, sample images were transformed by spatial warp of the $x$-axis onto the $z$-axis ($\psi_{xz}$), $y$-axis onto the $z$-axis ($\psi_{yz}$) and the $x$~\&~$y$-axes uniformly onto the $z$-axis ($\psi_{xz}\!=\!\psi_{yz}$). For each set, transformation coefficients were sampled at increments of 0.00005 over a range of 0.0 to 0.01 for a total of 201 samples per set. A select set of sampled transformations are shown in Table~\ref{table:Warp}, showing estimation by Equation~\ref{eq:dft_warp} for singular points. This table shows conformity with uniform scaling but poor handling of linearity due to estimation of singular points without spatial awareness, knowledge of neighboring pixels, or recursive transformation of the spatial range. This divergence between estimated and experimental results is pronounced for $\psi_{xz}=\psi_{yz}$ and occurs at later increments for $\psi_{xz}$ and $\psi_{yz}$ (not shown). The complete sampled range can be visualized in Figure~\ref{plots:WarpXY} where to allow visualization of the non-uniform scaling shown in Equation~\ref{eq:dft_warp} the solid lines represent values from the magnitude's maxima locations and the often overlapping dotted lines represent values from original encoded locations. 

Vertices in the solid lines correspond to the changing of coordinates for the magnitude's maxima, where at the first vertex the solid curves diverge from the values of original encoded locations (dotted lines). Vertices occur at a warp of 0.0026 in Figure~\ref{plot:WarpX_Phase} and 0.002, 0.0037, 0.0042, and 0.006 in Figure~\ref{plot:WarpXY_Phase}. From these plots, it can be observed that Cartesian quadrant pairs (quadrants mirrored in Fourier periodicity, I\&III, and II\&IV) often overlap and are impacted uniformly. The values and corresponding figures for spatial warping of the $y$-axis have been omitted for they were visually identical to spatial warping of the $x$-axis illustrated in Figures~\ref{plot:WarpX_Phase}-\ref{plot:EstWarpX_Phase}.

\section{Discussion}
\label{sec:Discussion}

Bracewell~et~al.'s Affine Transformation Theorem created the foundation for the extension to perspective transformation pairs. For affine transformations, the novel formulation presented here reduces to Bracewell's equation, though it should be noted that equations reported are for DFT and more closely match the observations of \textit{Fast Fourier Transform} and \textit{Fast Fourier Transform And Its Applications} by Brigham~\cite[~p.35]{book:fastFourierTransform}\cite[~p.30-47]{book:fastFourierTransformAndApplications}. While this work has described these equations to model 2D images in 3D space, there is no theoretical limitation to extending the dimensionality of these equations. The selected scope was chosen to facilitate simulation and model real-world applications, although the mathematics of SFTP has been solved for spatial perspective transformations of 3D models in 3D space. 

Through Section \ref{sec:experimentalResults} and Section \ref{sec:LimitationAndSpecialCases}, it has been demonstrated that the Fourier spectrum can be sampled discretely and continuously. At lower levels of the transformation hierarchy this periodic property of DFTs allows for selective resolution in pattern sampling and thus control over the computational load. Though transformation coordinates in this work were sampled as the maxima of normalized cross-correlation template matching, future work can increase resolution and enable sub-pixel precision by modeling the decomposition of encoded values and increasing DFT frequency resolution.

\begin{equation}
	C(u,v) = \alpha(u)\alpha(v)\cos\left(\frac{(2x+1)u\pi}{2*N}\right)\cos\left(\frac{(2y+1)v\pi}{2N}\right)
	\label{eq:dct}
\end{equation}

\begin{equation}
	\alpha(u) = 
	\begin{cases}
		\sqrt{\frac{1}{N}} &\text{ for }u=0\\
		\sqrt{\frac{2}{N}} &\text{ for }u=1,2,...,N{-}1\\
	\end{cases}
	;
	\alpha(v) = 
	\begin{cases}
	\sqrt{\frac{1}{N}} &\text{ for }v=0\\
	\sqrt{\frac{2}{N}} &\text{ for }v=1,2,...,N{-}1\\
	\end{cases}
\end{equation}

Future work includes potential pursuit of SFTP with other frequency domain algorithms such as Discrete Cosine Transform (DCT) to provide faster computational performance and more succinct results. As described by Gonzalez~et~al.~\cite[475-478]{book:digimproc4ed}, there are few advantages and disadvantages to the implementation of a DCT over a DFT. Similar to DFTs, the computational time of a DCT is non-monotonic, so array sizes should be factor-able as a product of small prime numbers or padded for processing efficiency~\cite{web:opencvDocs}. As documented by OpenCV~\cite{web:opencvDocs} and described by Gonzalez~et~al.~\cite[~p.477]{book:digimproc4ed}, 2D DCTs have a restriction to the resolution (periodicity) which forces the resolution of an $M\!\times\!N$ DCT to the size of one dimension. This is observable in Equation~\ref{eq:dct} where the periodicity of the $u$ and $v$-axes of a DCT are set to $2N$.

DCTs are more computationally efficient than DFTs, largely as a result of being real-valued rather than complex-valued like DFTs. Though the frequency ranges expressed by DCTs and DFTs are almost the same, DCTs have double the frequency resolution of DFTs. This is a result of DFTs reflecting across the ZF, cutting the maximum number of frequencies in half. DCTs assume symmetry, meaning that the DCT of an $N$-point function $f(x)$ can be obtained from the DFT of a $2N$-point symmetrically extended version of $f(x)$. This feature of DCTs minimizes boundary discontinuities which can introduce high-frequency components in DFTs. Overall, DCTs share many attributes with DFTs but set different assumptions and are more computationally efficient.

With respect to applications of this work, it is recognized that for any given encoded pattern, projected onto a flat incident surface, the DFT of the pattern can be utilized to interpret the uniform transformation of the encoded pattern from projector space to camera space. This is feasible as the transformation is uniform across the entirety of the captured image which provides a distinct Euclidean, similarity, affine or perspective transformation. For non-planar surfaces, the transformation of any point can be described on the tangent plane as a linear transformation by a localized DFT. Using these principles, future work will explore the potential for spatio-temporal Fourier surface imaging sensors to recover surface normals. 

\section{Conclusions}
\label{sec:Conclusion}

In this work the mathematical relationship for Spatial-Fourier Transformation Pairs (SFTP) was derived, defining the transformation of spatial transformed planar surfaces in the Discrete Fourier Transform (DFT) spectrum. The mathematical relationship for the twelve degrees of freedom in perspective transformation was defined, expanding on previously derived affine Spatial-Fourier Transformation Pairs (SFTP). The Perspective Transformation Theorem was proposed as an extension to Bracewell~et al.'s Affine Transformation Theorem. The theorem was validated experimentally, by transforming sample images in the spatial domain and analyzing them in Fourier space to assert congruity with independent and uniform transform pairs for Euclidean, similarity, affine and ultimately perspective transformations.

\section*{Funding}
Epson Canada, Limited; Ontario Centers of Excellence (VIP II 27531); Natural Sciences and Engineering Research Council of Canada (NSERC) (CRDPJ506235-16).

\section*{Acknowledgments}
The authors would like to acknowledge Epson Canada, the Natural Sciences and Engineering Research Council of Canada, and the Ontario Centres of Excellence, for their support of this work.

\section*{Disclosures}

\medskip

\noindent The authors declare no conflicts of interest.


\bibliography{Bibliography}

\end{document}